%% file: main.tex
\documentclass[aps,prl,reprint,twocolumn,superscriptaddress,nofootinbib,floatfix,raggedbottom]{revtex4-2}

\input{preamble.tex}

\begin{document}

\title{Quantum measure transport for universal sampling and generative artificial intelligence}

\author{Benjamin Savinson\,\textsuperscript{*}}
\affiliation{Optical Materials Engineering Laboratory, Department of Mechanical and Process Engineering, ETH Zurich, 8092 Zurich, Switzerland.}
\affiliation{Seminar for Applied Mathematics, Department of Mathematics, ETH Zurich, 8092 Zurich, Switzerland.}
\affiliation{ETH AI Center, ETH Zurich, 8092 Zurich, Switzerland.}

\author{Andrey Bryutkin\,\textsuperscript{*}}
\affiliation{Department of Mathematics, Massachusetts Institute of Technology, Cambridge, Massachusetts 02139, USA.}
\affiliation{Laboratory for Information and Decision Systems, Massachusetts Institute of Technology, Cambridge, Massachusetts 02139, USA.}

\author{Oliver R.\ M\"uller}
\affiliation{Optical Materials Engineering Laboratory, Department of Mechanical and Process Engineering, ETH Zurich, 8092 Zurich, Switzerland.}

\author{Boris de Jong}
\affiliation{Debye Institute for Nanomaterials Science, Utrecht University, Utrecht 3584CC, The Netherlands.}

\author{Tjom Arens}
\affiliation{Debye Institute for Nanomaterials Science, Utrecht University, Utrecht 3584CC, The Netherlands.}

\author{Siddhartha Mishra}
\affiliation{Seminar for Applied Mathematics, Department of Mathematics, ETH Zurich, 8092 Zurich, Switzerland.}
\affiliation{ETH AI Center, ETH Zurich, 8092 Zurich, Switzerland.}

\author{Sander J.\ W.\ Vonk}
\affiliation{Optical Materials Engineering Laboratory, Department of Mechanical and Process Engineering, ETH Zurich, 8092 Zurich, Switzerland.}

\author{Youssef Marzouk\textsuperscript{\ensuremath{\dagger}}}

\affiliation{Laboratory for Information and Decision Systems, Massachusetts Institute of Technology, Cambridge, Massachusetts 02139, USA.}

\author{David J.\ Norris\textsuperscript{\ensuremath{\ddagger}}}

\affiliation{Optical Materials Engineering Laboratory, Department of Mechanical and Process Engineering, ETH Zurich, 8092 Zurich, Switzerland.}
\affiliation{ETH AI Center, ETH Zurich, 8092 Zurich, Switzerland.}

\begin{abstract}
\sloppy

Generative artificial intelligence requires samples drawn from complex, high-dimensional probability distributions. Digital computers map pseudo-random samples from a simple reference (such as a Gaussian) to a desired target---a process known as measure transport. While state-of-the-art methods can learn these mappings with high fidelity, the underlying transformations require many model evaluations or numerical integration steps, increasing cost. Here we show that quantum systems governed by the Schrödinger equation can perform measure transport and sampling directly in physical hardware. Using appropriate potentials, particle wavefunctions are guided so that measurement statistics approximate prescribed target distributions, with individual measurements providing samples.  We prove that such systems can represent arbitrary probability distributions, establish a formal equivalence to flow matching, and show how triangular decompositions enable high-dimensional sampling in low-dimensional physical systems. For experiments, we exploit single-photon detection to generate images and text, establishing a path toward highly parallel, energy-efficient sampling for real-world statistical models.

\end{abstract}

\maketitle
\begingroup
\renewcommand{\thefootnote}{\ensuremath{\dagger}}
\footnotetext{ymarz@mit.edu}
\renewcommand{\thefootnote}{\ensuremath{\ddagger}}
\footnotetext{dnorris@ethz.ch}
\endgroup
\linespread{0.97}\selectfont

\input{extended_data.tex}

Efficient sampling from high-dimensional probability distributions is a key challenge in many fields, including statistical physics\cite{Faulkner2024}, computational chemistry\cite{Cheng2025, Zhu2026}, and artificial intelligence (AI)\cite{Papamakarios2021}. Modern approaches, such as diffusion models\cite{Song2021} and flow matching\cite{Lipman2023,Albergo2025}, use measure transport\cite{Marzouk2016} to map simple reference distributions to complex target distributions. In practice, these models typically require many sequential computational steps. Diffusion samplers progressively remove noise\cite{Song2021} and flow-matching samplers numerically integrate a learned vector field\cite{Lipman2023, Albergo2025}. Although accelerated sampling methods can reduce the number of steps\cite{Salimans2022}, substantial computational costs remain when generating many samples. Consequently, sampling from multimodal and high-dimensional distributions remains computationally demanding and difficult to scale.

\begingroup\looseness=-1
Physical systems provide a potential route to more efficient computations\cite{Yasuda2021, Zangeneh-Nejad2021}.  In particular, photonic devices offer low loss, reduced latency, and intrinsic parallelism\cite{Athale2016, Lin2018, Wetzstein2020, McMahon2023}. This has motivated the use of light to perform matrix--vector multiplication\cite{Yang2013, Nikkhah2024} and nonlinear transformations\cite{Momeni2022, Eliezer2023, Momeni2023, Wanjura2024, Xia2024, Yildirim2024} relevant to AI workloads\cite{Kalinin2025}. In addition, stochastic computations have been performed by digitally generating random initial conditions and using optical processors to transform them toward target distributions\cite{Oguz2024, Chen2025}. While this shifts part of the process to the optical domain, the randomness (actually pseudorandomness) is still generated digitally, and the latent variables or intermediate states must be encoded in the optical fields. Thus, the overall sampling pipeline remains constrained by digital-control and data-conversion overheads.

\par\endgroup

An alternative approach is to utilize the intrinsic randomness of physical measurements for sampling. For example, single photons propagating through optical interferometers can be used to draw samples from discrete probability distributions defined by the device\cite{Reck1994}. Additional sources of randomness such as phase noise\cite{Lovic2021} and intensity fluctuations can also be exploited\cite{Bruckerhoff2025}. Further, intrinsic optical randomness has enabled stochastic neural-network inference\cite{Ma2025} and autoregressive image generation\cite{Choi2024}. However, these approaches rely on device-specific stochastic responses and do not establish a general framework for transporting an initial probability density to a target distribution prescribed by data. Unifying this data-driven measure transport with sampling in a single programmable physical process remains an open challenge. Here, we present quantum measure transport (QMT) as a general framework to address arbitrary target probability distributions with quantum systems. By designing the potential in Schrödinger’s equation, wavefunction propagation and detection allow efficient measure transport and sampling. QMT can also be trained directly on samples, without additional information about the target distribution.

\begin{figure*}[t]
    \centering
    \includegraphics[width=\linewidth]{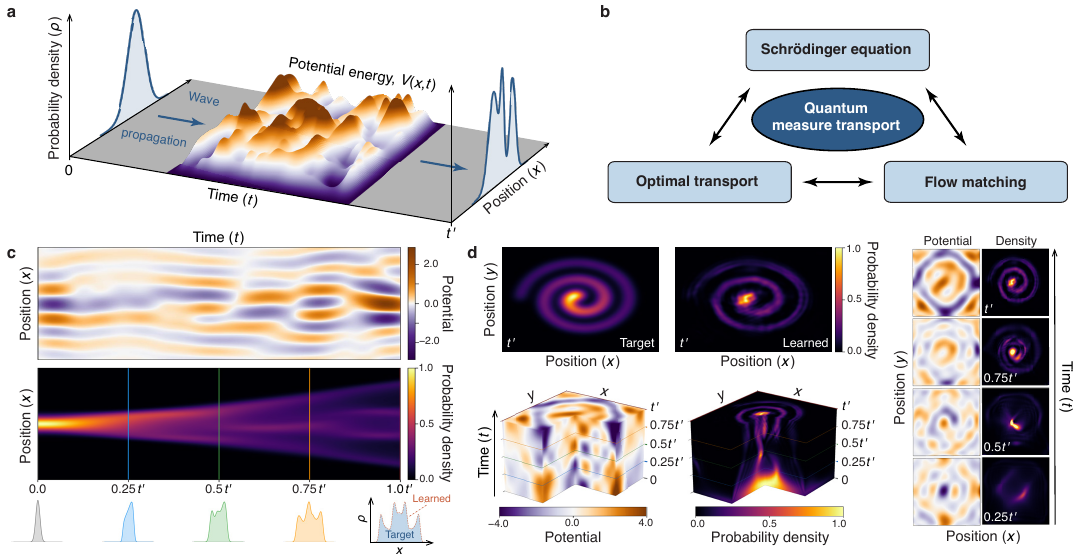}
    \caption{\textbf{Quantum measure transport (QMT).} \textbf{a}, Concept of QMT: a quantum-mechanical particle with a simple initial probability density $\rho$ evolves under the Schr\"odinger equation through an engineered time-dependent potential-energy function $V(x,t)$ into a prescribed target density $\rho_{\text{tar}}(x)$ at time $t'$. \textbf{b}, QMT connects Schr\"odinger evolution, optimal transport, and flow matching; the learned potential plays the role of the flow-matching velocity field. \textbf{c}, One-dimensional QMT: a learned potential $V(x,t)$ (top) results in the density evolution (middle). Vertical slices (bottom) show the quantum evolution of the probability densities at $t=$~0.0, 0.25\,$t'$, 0.5\,$t'$, 0.75\,$t'$, and 1.0\,$t'$. At $t'$, the learned density (dashed line) is compared to the target (shaded region). \textbf{d}, Two-dimensional QMT for a spiral target. Target and learned densities at $t'$ (top); volumetric renderings of $V(x,y,t)$ and $\rho(x,y,t)$ (bottom left); transverse slices in $t$ of both $V$ and $\rho$ (right). All probability densities are normalized to their maximum value.}
    \label{fig:1}
\end{figure*}

\section{Principles of quantum measure transport}

A quantum-mechanical particle that propagates through a one-dimensional (1D), time-dependent potential-energy function $V(x,t)$ is described by the Schr\"odinger equation
\begin{equation}
    i\hbar \frac{\partial}{\partial t} \psi(x,t) = \left\{-\frac{\hbar^2}{2m}\frac{\partial^2}{\partial x^2} + V(x,t) \right\}\psi(x,t)~,
\label{eq: schrodinger}
\end{equation}
with $\psi$ the particle wavefunction, $m$ its mass, and $\hbar$ the reduced Planck constant\cite{Griffiths}. At any time $t$, the probability density $\rho$ of detecting the particle at position $x$ is $|\psi(x,t)|^2$.   The goal of QMT (Fig.~\ref{fig:1}a) is to guide the particle from an initial distribution $|\psi(x,0)|^2$ at $t=0$ to a target distribution $\rho_{\text{tar}}(x)=|\psi(x,t')|^2$ at a later $t'$. In this case, particle propagation performs the measure transport while particle detection at $t'$ provides the sampling. To implement this, a potential $V_{\text{tar}}(x,t)$ that evolves the wavefunction as required must be identified.

We prove the existence of such a potential $V_{\text{tar}}(x,t)$ for any prescribed $\rho_{\text{tar}}(x)$ using results from optimal transport theory (Supplementary Notes 1–2). We further show  the equivalence between QMT and flow matching in generative modelling\cite{Lipman2023} (Supplementary Note 1). The velocity field learned by a flow-matching model maps directly to the quantum potential  governing the Schrödinger evolution, placing QMT on the same theoretical footing as state-of-the-art generative AI. The relationships between these concepts are summarized in Fig.~\ref{fig:1}b.

To determine $V_{\text{tar}}(x,t)$, we deploy gradient-based inverse design\cite{Leung2017}. The target density $\rho_{\text{tar}}$, either given in closed form or estimated from samples, defines a loss function that measures the discrepancy between the evolved density $|\psi(x,t')|^2$ and $\rho_{\text{tar}}$ (Methods). We discretize equation \eqref{eq: schrodinger} using a unitary time-integration scheme that preserves normalization of the wavefunction throughout the evolution, and we obtain gradients of the potential via automatic differentiation. Figure~\ref{fig:1}c shows the learned potential and distribution for a representative 1D target distribution, demonstrating accurate recovery of a multimodal $\rho_{\text{tar}}$ from data samples. By solving equation (1) in two spatial dimensions, this scheme can be extended to two-dimensional (2D) probability distributions. For example, Fig.~\ref{fig:1}d presents the learned potential $V_{\text{tar}}(x,y,t)$ that maps a 2D Gaussian to a target spiral distribution.

\begin{figure*}[t]
    \centering
    \includegraphics[width=\linewidth]{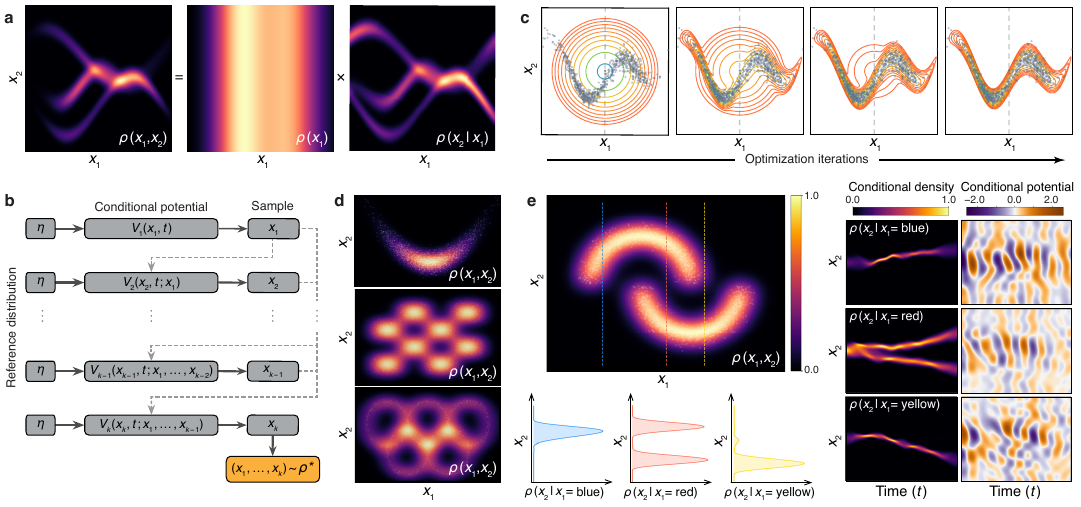}
    \caption{\textbf{High-dimensional sampling via triangular transport.} \textbf{a}, Decomposition of a two-dimensional target probability density $\rho(x_1, x_2)$ into the marginal $\rho(x_1)$ and  conditional $\rho(x_2\!\mid\!x_1)$ probability densities. Each plot is independently normalized for visualization. \textbf{b}, Sequential sampling: for each dimension $k$, the conditional potential $V_k(x_k,t;x_1, \dots, x_{k-1})$ evolves the reference $\eta$ and yields $x_k$, which conditions all subsequent potentials (dashed lines). The full vector represents $\rho^*$, an estimate of the target distribution $\rho$. \textbf{c}, Reference samples and isodensity contours warping onto the data points over 0, 10, 30, and 100 optimization iterations. The contours represent 1.8\%, 2.8\%, 4.3\%, 6.7\%, 10\%, 16\%, 25\%, 38\%, 59\%, and 92\% of the maximum density. \textbf{d}, Learned two-dimensional probability densities for banana, checkerboard, and rings (Methods). \textbf{e}, Double crescent target with conditional slices $\rho(x_2\!\mid\!x_1)$ at the blue, red, and yellow vertical lines. On the right, conditional probability densities are plotted with the corresponding learned potentials for the three vertical lines. The density maps in \textbf{d} are normalized using the scale in \textbf{e}.} 
    \label{fig:2}
\end{figure*}

\begin{figure*}[t]
    \centering
    \includegraphics[width=\linewidth]{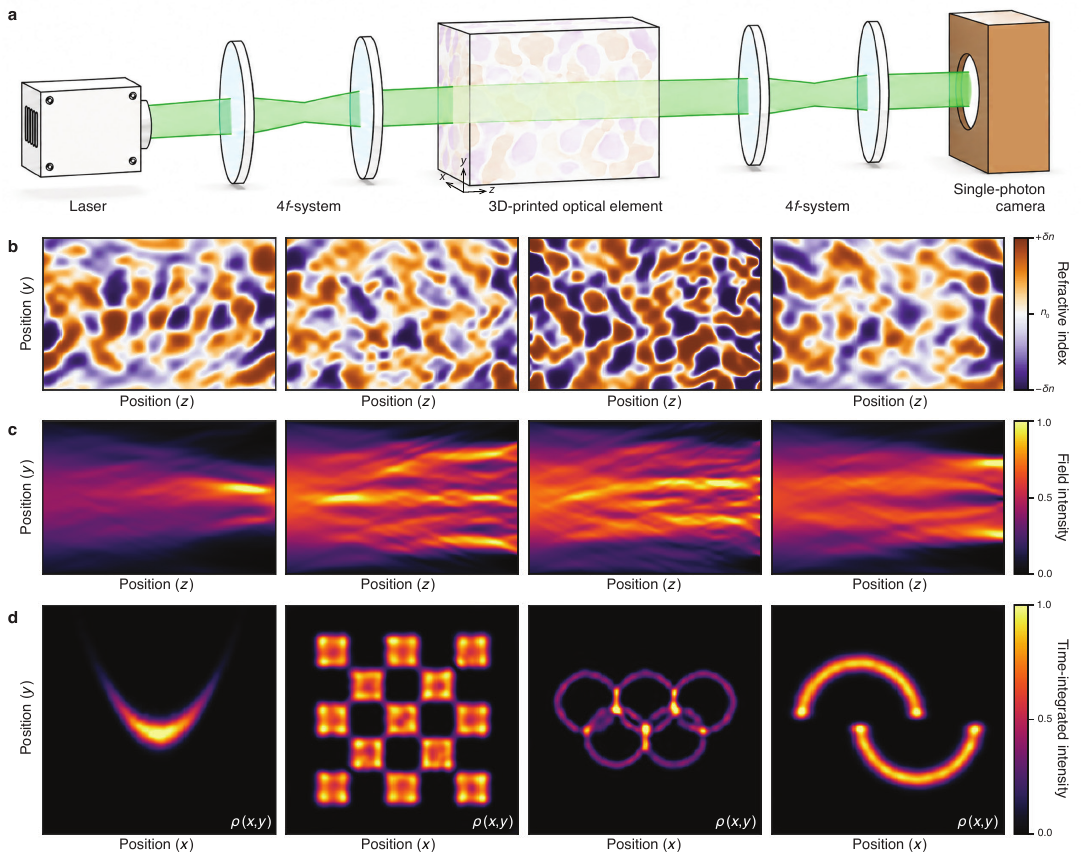}
    \caption{\textbf{Simulations of photonic QMT.} \textbf{a}, Schematic of an apparatus for photonic QMT. A collimated beam with $\lambda_0=1550\,\mathrm{nm}$ propagates in $z$ through an inverse-designed block ($110\times110\times220\,\upmu\mathrm{m}^3$) with a refractive index $n(x,y,z)$ that spatially varies between 1.5113 and 1.5213 ($n_{\text{ref}} =$ 1.5163 and $\pm\delta n=\pm$0.0050). These values are typical of optical glasses and polymers, and the index contrast is well within the range reported for 3D-printed gradient-index optics (refs. \citenum{Dylla2020}, \citenum{Gao2023}). The block transforms the beam into the target distribution, and the output plane is imaged onto a single-photon detector array, where each detection event constitutes one sample. \textbf{b}, Optimized refractive-indices $n(x,y,z)$ to obtain four different target probability distributions (banana, checkerboard, rings, and double crescent; columns). The plots show the $yz$ plane at $x=$\,40\,$\upmu\mathrm{m}$, in the middle of the block. \textbf{c}, Corresponding simulated intensity evolution through the block (for the same plane as in \textbf{b}). Propagation is from left to right. \textbf{d}, Simulated output intensity at the detector plane, showing the target probability densities. }
    \label{fig:3}
\end{figure*}

\section{High-dimensional sampling via triangular transport}
\label{sec: high_dim}

Real-world sampling problems often involve probability distributions in very high-dimensional spaces, far exceeding the physical dimensionality of any single quantum system. To address this, we combine QMT with triangular transport maps\cite{Ramgraber2025}. Any $D$-dimensional target probability density can be decomposed into a product of 1D conditionals,
\begin{equation}
\begin{aligned}
\rho(x_1,\dots,x_D)
={}&\rho(x_1)\,\rho(x_2\!\mid\!x_1)\\
&\quad\rho(x_3\!\mid\!x_1,x_2)\dots\\
&\quad\rho(x_D\!\mid\!x_1,\dots,x_{D-1})~,
\end{aligned}
\label{eq:chain_rule}
\end{equation}
where $\rho(x_k\!\mid\!x_1, \dots, x_{k-1})$ is the conditional probability density of $x_k$ given $x_1, \dots, x_{k-1}$ (Supplementary Note 3).  Figure~\ref{fig:2}a shows an example of this decomposition for a 2D distribution.  QMT realizes such decompositions as a sequence of 1D quantum evolutions. For each dimension $k$, a conditional potential $V_k(x_k, t; x_1, \dots, x_{k-1})$ is designed to produce samples from $\rho(x_k\!\mid\!x_1, \dots, x_{k-1})$, with the previously sampled variables $x_1, \dots, x_{k-1}$ as external control biases that modulate the potential (Fig.~\ref{fig:2}b).

Therefore, in contrast to the potential $V_{\text{tar}}(x,y,t)$ shown in Fig.~\ref{fig:1}d, the conditional potentials $V_k$ change as a function of the previous measurements. To capture this relationship, we use a multi-layer perceptron (MLP) that takes $x_1, \dots, x_{k-1}$ as conditioning inputs and yields the potential $V_k$ for the $k$-th evolution (Methods). The MLP learns the required mapping during training by solving equation \eqref{eq: schrodinger} for the predicted potential and iteratively minimizing the Kullback--Leibler (KL) divergence\cite{Cover2006} to a target distribution. Figure~\ref{fig:2}c demonstrates how an initial Gaussian distribution is transformed to match a set of target samples as a function of optimization iterations using conditional potentials. We validate this approach on several 2D target distributions (Fig.~\ref{fig:2}d), including multimodal and correlated densities. Figure~\ref{fig:2}e shows how the 1D quantum evolutions used to sample the second variable $x_2$ are conditioned on the results of the first measurement $x_1$. These results illustrate how triangular transport enables sampling in higher dimensions without modifying the dimensionality of the underlying quantum system. It relies only on the expressivity of the learned conditional potentials.

\begin{figure*}[t]
    \centering
    \includegraphics[width=\linewidth]{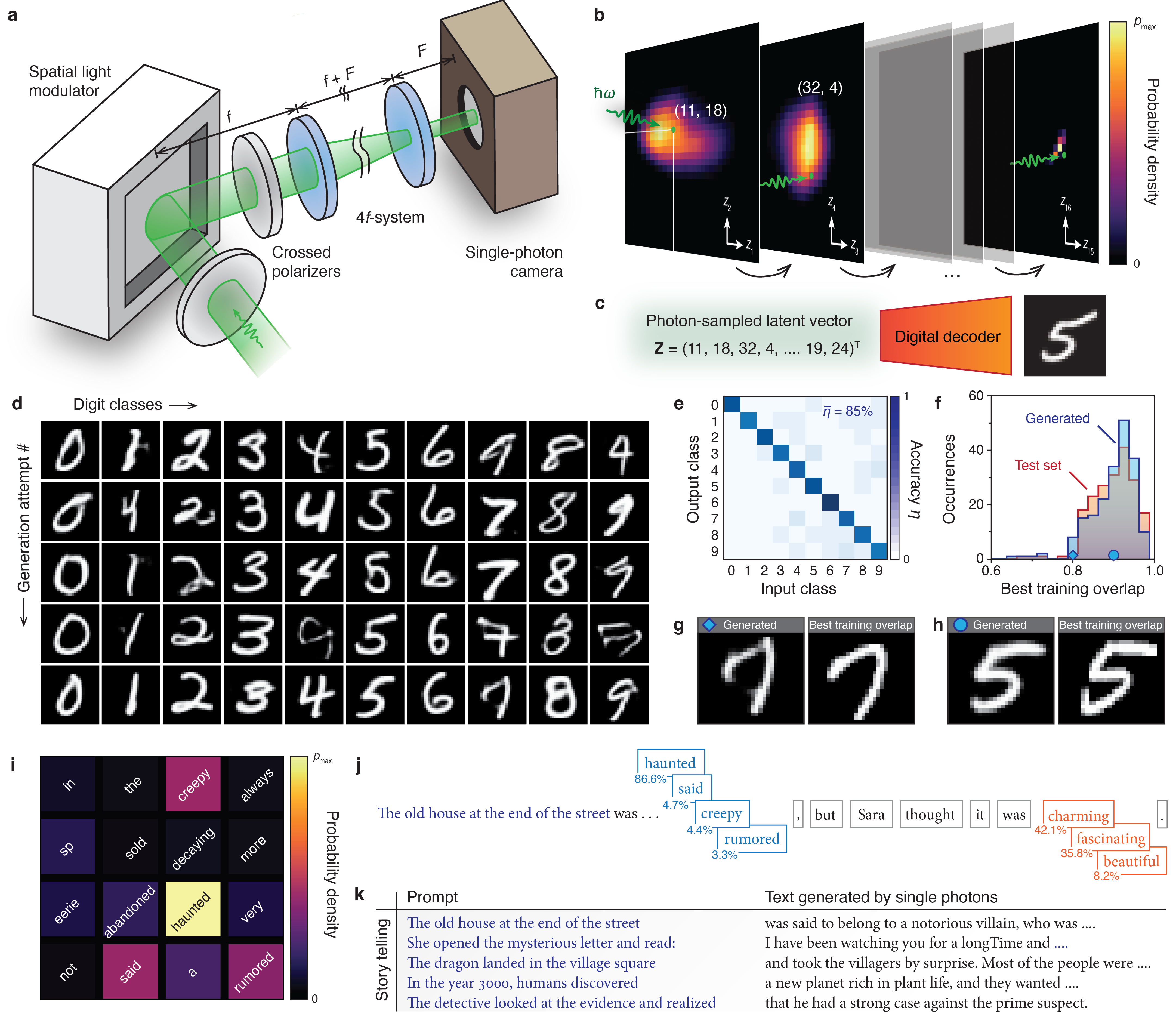}
    \caption{\textbf{Single-photon sampling for image and text generation.} \textbf{a}, Schematic of the single-photon experimental setup for sampling. A spatial light modulator placed between crossed polarizers modulates the amplitude profile, encoding the target probability distribution, which is imaged onto a single-photon camera. At low light intensity,  single-photon frames can be used for sampling. \textbf{b}, Image generation follows a sequence of 2D conditional distributions, each conditioned on the previous sample, learned from autoencoder-compressed images using triangular transport maps. The starting distribution is fixed and determined by the desired digit class. Each single-photon sample yields two numbers; eight sequential samples fill a 16-dimensional latent vector. \textbf{c}, A digital decoder transforms the constructed latent vector into a new image. \textbf{d}, The first five generation attempts per digit class from the sampling experiment. \textbf{e}, Confusion matrix for all experimentally generated digits, classified by a digital neural network (accuracy $\eta_{\mathrm{test}}$=97\%). The average classification accuracy for generated digits is $\eta_{\mathrm{gen}}=$\,85\%. \textbf{f}, Histogram of best overlap (cosine similarity) with items in the training set, indicating that the experimentally generated digits (blue) are not direct copies of the training set. The same analysis between the test and training set shows similar best overlaps (red); overlapping bars appear grey. \textbf{g}, Selected pair of generated/training digits with 80\% best overlap, showing similar features but clear differences. \textbf{h}, Same as \textbf{g}, but for a digit pair with 90\% overlap. \textbf{i}, For optical text generation, the possible next tokens are mapped to distinct pixel regions (7×7 pixels per token) on the single-photon camera. Token probabilities are generated by the Microsoft Phi-2 large language model (temperature $T=1$). \textbf{j}, Example for a storytelling prompt, indicating the first 10 next tokens and highlighted alternative probabilities for intermediate tokens. \textbf{k}, Experimentally generated text for five different prompts. Extended Data Figs.~\ref{EDfig:3} and \ref{EDfig:4} show further experimentally generated digits and texts.}
    \label{fig:4}
\end{figure*}

\section{Photonic implementation}

While different quantum-mechanical particles could be used for QMT, we select photons. In addition to the advantages already mentioned for light, single photons can be easily detected. We can also exploit the equivalence between the Schr\"odinger equation and the paraxial wave equation for scalar electromagnetic fields\cite{Milonni1992}. Namely, assuming only low-angle scattering and negligible back-reflections, the electric-field envelope $E(x,z)$ for light travelling in the $z$-direction through a medium with a spatially varying refractive index $n(x,z)$ is (Supplementary Note 4)
\begin{equation}
\begin{aligned}
i\frac{\partial}{\partial z}E(x,z)
={}&\Bigl\{-\frac{1}{2k_0 n_{\text{ref}}}
\frac{\partial^2}{\partial x^2}\\
&-\frac{k_0}{2n_{\text{ref}}}
\bigl[n^2(x,z)-n_{\text{ref}}^2\bigr]\Bigr\}
E(x,z)~,
\end{aligned}
\label{eq:optical_schrodinger}
\end{equation}
\noindent where $k_0 = 2\pi/\lambda_0$ with $\lambda_0$ the optical wavelength in vacuum, and $n_{\text{ref}}$ is the reference refractive index. This expression is identical to equation \eqref{eq: schrodinger}, with the propagation direction $z$ replacing $t$ and the second term in brackets (related to the refractive index) replacing the potential $V$. The intensity profile $I(x, z)=|E(x, z)|^2$ represents the probability density. Thus, QMT can be performed by sending light with an initial intensity profile $I(x, z=0)$ through a medium with a designed refractive-index landscape. By detecting the position $x$ of each photon at $z'$, the target distribution $I(x, z')$ is sampled. 

We simulate this approach by extending equation~\eqref{eq:optical_schrodinger} to a beam with two transverse dimensions, $x$ and $y$. We consider an apparatus (Fig.~\ref{fig:3}a) in which photons propagate in $z$ through a block ($110\times110\times220\,\upmu\mathrm{m}^3$) with a refractive index $n(x,y,z)$ that spatially varies between 1.51 and 1.52\nocite{Dylla2020,Gao2023}. We assume a camera with single-photon sensitivity detects where each transmitted photon lands, yielding $I(x, y, z')$. If $n(x,y,z)$ is properly designed, this leads to sampling from a 2D target distribution. Figure~\ref{fig:3}b,c plots examples of $n$ and the electric field for the target probability distributions shown in Fig.~\ref{fig:3}d.

\section{QMT experiments}

To combine such an apparatus with triangular transport, the block in Fig.~\ref{fig:3}a must be reconfigurable. The index $n(x,y,z)$, representing the conditional potentials $V_k$, must change after every sample. While photonic devices exist\cite{Onodera2026} that could implement such 1D conditional potentials, we demonstrate QMT sampling experimentally with a 2D detector, so that each detected photon samples a pair of variables $(x_k, x_{k+1})$ from $\rho(x_k, x_{k+1}\!\mid\!x_1, \dots, x_{k-1})$ ($k$ odd). For this purpose, we replace the block with a spatial light modulator (SLM) (Fig.~\ref{fig:4}a and Extended Data Fig.~\ref{EDfig:1}a). Placed between crossed polarizers and imaged onto a single-photon camera, the SLM collapses the block into a single layer with potential $V(x,y,z)=V(x,y)\,\delta(z)$ (Supplementary Note 5). This provides reconfigurability at the cost of physical compactness and photon efficiency. The block with its three-dimensional (3D) potential can generate targets that a 2D potential must resolve pixel by pixel.

Using our setup, we perform single-photon QMT and demonstrate two examples of generative AI: images and text. For image generation, an autoencoder was trained digitally to compress images of MNIST handwritten digits into a 16-dimensional latent space, and the conditional densities required for triangular transport were estimated from the training data (Methods). Sampling from the learned latent distribution was then done physically. For each pair of latent coordinates, the SLM encoded the conditional density $\rho(x_k, x_{k+1}\!\mid\!x_1, \dots, x_{k-1})$ and a single-photon detection event drew both coordinates (Fig.~\ref{fig:4}b,c and Extended Data Fig.~\ref{EDfig:2}a). The complete latent vector was decoded digitally to produce the output image. Figure~\ref{fig:4}d shows generated digits across all ten classes over multiple generation attempts. The confusion matrix (Fig.~\ref{fig:4}e) confirms high class fidelity, and a training-overlap analysis (Fig.~\ref{fig:4}f--h) establishes that generated images are novel rather than reproductions of training examples. Further images are presented in Extended Data Fig.~\ref{EDfig:3}.

For text generation, we performed next-token selection with a large language model. Next-token probabilities for a given prompt were computed digitally using a pre-trained language model\cite{Javaheripi2023} and mapped to pixels on the SLM, with optical intensity set proportional to token probability (Fig.~\ref{fig:4}i; Methods). A single-photon detection event selects the next token according to this distribution, with the process repeated autoregressively to generate text. Figure~\ref{fig:4}j--k and Extended Data Fig.~\ref{EDfig:2}b show next-token distributions and generated text for several story-telling prompts, demonstrating coherent multi-token generation driven by single-photon sampling events. Additional data are given in Extended Data Fig.~\ref{EDfig:4}.

\section{Discussion and outlook}

We have shown that single-particle quantum systems perform measure transport and sampling. Practical applications of QMT must involve high-dimensional distributions as well as fast reconfigurability and high-resolution detection. In photonic systems, the detection rate and resolution are primarily set by the tunable optical element and the single-photon detector. Integrated photonic circuits with electro-optic phase shifters could implement 1D conditional potentials at gigahertz rates in compact footprints\cite{Wang2018}. Further, switchable elements, in which a control beam writes the refractive-index profile, would allow the potential to be optically reconfigured\cite{Onodera2026}. For the detector, superconducting nanowire arrays could provide high-efficiency detection and low dark counts, particularly at telecom wavelengths\cite{Oripov2023}.

We addressed the need for high-dimensional sampling with triangular decomposition. However, this approach also introduces trade-offs. In particular, the conditional potentials become expensive to compute as the dimension grows. Fortunately, this expense can be mitigated 
by exploiting the sparsity structure of triangular measure transport\cite{Spantini2018}. Further, the physical dimensionality of the quantum system itself could be increased. Entangled photons would yield correlated detection events. A single pass could then sample joint distributions over several variables at once. Further, interacting bosons, such as ultracold atoms in optical lattices, would offer an additional route to introduce coupling between different particles. Thus, quantum measure transport provides many routes to explore physics-based hardware for generative AI and other high-dimensional statistical models using quantum systems.

\input{methods}

\begin{addendum}
\vspace{12pt}
 \item [Acknowledgements] We thank F.~Rabouw for technical support and P.~Tsimpos for helpful discussions on universality. This work was supported by ETH Zurich (Research Grant 25-1 ETH-030), the Swiss National Science Foundation (SNSF; Award No.~2000-1-240090), the US Department of Energy (DOE) SciDAC-5 program (contract DE-SC0012704, subcontract 425236), and the DOE Office of Advanced Scientific Computing Research (award DE-SC0023187).
 
 \item[Author contributions] B.S., A.B., Y.M., and D.J.N. conceived the project with input from S.M.\quad B.S.\ and A.B. developed the theory behind QMT with input from Y.M.\quad B.S.\ and A.B. performed the inverse design of the potentials and the electromagnetic simulations. The experiments were designed by B.S., O.M., S.J.W.V., and D.J.N.\ and performed by O.M., B.d.J., T.A., and S.J.W.V.\quad  B.S., A.B., Y.M., and D.J.N.\ wrote the manuscript with input from all authors.\quad D.J.N., S.M., and Y.M.\ supervised the project.

 \item[Competing Interests] The authors declare the following potential competing financial interests: B.S., A.B., Y.M., and D.J.N. have filed a patent application related to ideas in this work. 
 
 \item[Additional information] 
 \leavevmode\\[0.5ex] 
 \textbf{Correspondence and requests for materials} should be addressed to Y.M. or D.J.N.
\end{addendum}

\setlength{\bibsep}{0pt}
\renewcommand{\bibfont}{\small}
\bibliography{refs}

\clearpage
\EDFigureOne
\EDFigureTwo
\EDFigureThree
\EDFigureFour

\end{document}


\title{\texorpdfstring{Supplementary information for\\ `Quantum measure transport for universal sampling and generative artificial intelligence'}{Supplementary information for 'Quantum measure transport for universal sampling and generative artificial intelligence'}}
\author{Benjamin Savinson}
\thanks{These authors contributed equally.}
\affiliation{Optical Materials Engineering Laboratory, \vspace{-0.7em} \\
Department of Mechanical and Process Engineering,\vspace{-0.7em} \\
ETH Zurich, 8092 Zurich, Switzerland.}
\affiliation{Seminar for Applied Mathematics, \vspace{-0.7em} \\
Department of Mathematics, \vspace{-0.7em} \\
ETH Zurich, 8092 Zurich, Switzerland.}
\affiliation{ETH AI Center, \vspace{-0.7em} \\
ETH Zurich, 8092 Zurich, Switzerland.}
\author{Andrey Bryutkin}
\thanks{These authors contributed equally.}
\affiliation{Department of Mathematics, \vspace{-0.7em} \\
Massachusetts Institute of Technology, \vspace{-0.7em} \\
Cambridge, Massachusetts 02139, USA.}
\affiliation{Uncertainty Quantification Group, \vspace{-0.7em} \\
Laboratory for Information and Decision Systems, \vspace{-0.7em} \\
Massachusetts Institute of Technology, \vspace{-0.7em} \\ Cambridge, Massachusetts 02139, USA.}
\author{\\Oliver R. M\"uller}
\affiliation{Optical Materials Engineering Laboratory, \vspace{-0.7em} \\
Department of Mechanical and Process Engineering,\vspace{-0.7em} \\
ETH Zurich, 8092 Zurich, Switzerland.}
\author{Boris de Jong}
\affiliation{Debye Institute for Nanomaterials Science, \vspace{-0.7em} \\
Utrecht University, Utrecht 3584CC, The Netherlands.}
\author{Tjom Arens}
\affiliation{Debye Institute for Nanomaterials Science, \vspace{-0.7em} \\
Utrecht University, Utrecht 3584CC, The Netherlands.}
\author{Siddhartha Mishra}
\affiliation{Seminar for Applied Mathematics, \vspace{-0.7em} \\
Department of Mathematics, \vspace{-0.7em} \\
ETH Zurich, 8092 Zurich, Switzerland.}
\affiliation{ETH AI Center, \vspace{-0.7em} \\
ETH Zurich, 8092 Zurich, Switzerland.}
\author{\\Sander J.\ W.\ Vonk}
\affiliation{Optical Materials Engineering Laboratory, \vspace{-0.7em} \\
Department of Mechanical and Process Engineering,\vspace{-0.7em} \\
ETH Zurich, 8092 Zurich, Switzerland.}
\author{Youssef Marzouk}
\thanks{e-mail: ymarz@mit.edu}
\affiliation{Uncertainty Quantification Group, \vspace{-0.7em} \\
Laboratory for Information and Decision Systems, \vspace{-0.7em} \\
Massachusetts Institute of Technology, \vspace{-0.7em} \\ Cambridge, Massachusetts 02139, USA.}
\author{David J.\ Norris}
\thanks{e-mail: dnorris@ethz.ch}
\affiliation{Optical Materials Engineering Laboratory, \vspace{-0.7em} \\
Department of Mechanical and Process Engineering,\vspace{-0.7em} \\
ETH Zurich, 8092 Zurich, Switzerland.}
\affiliation{ETH AI Center, \vspace{-0.7em} \\
ETH Zurich, 8092 Zurich, Switzerland.}

\maketitle
\thispagestyle{empty}
\clearpage
\tableofcontents
\clearpage

\section{Hydrodynamic formulation of QMT and relation to flow matching}
\label{sec:si-theory}

This note gives the theory behind the main-text section `Principles of quantum measure transport'. We write the Schr\"odinger equation in Madelung variables (density and phase), show that a prescribed density evolution determines the potential, and relate QMT to flow matching (main-text Fig.~1b). Supplementary Note~2 uses these results to prove that QMT reaches every smooth, strictly positive density exactly and approximates every other distribution arbitrarily well.

\subsection{Periodic domain and absorbing boundaries}
\label{sec:si-setting}

Unless stated otherwise, Supplementary Notes~1 and~2 work on the periodic box $\Omega=[-L/2,L/2)^d$ with $d=1$ or $2$, that is, on the flat torus $\mathbb{T}^d_L=\R^d/(L\mathbb{Z})^d$. Probability that leaves the box through one face re-enters through the opposite face, so no further boundary conditions are needed. As a result, the Schr\"odinger evolution conserves probability (Lemma~\ref{lem:si-uniqueness}), and the equation that determines the phase has a unique solution up to a constant (Lemma~\ref{lem:si-phase-recovery}).

\begin{remark}[Other domains]\label{rem:si-domains}
On a bounded domain, the constructed wavefunction must also satisfy the boundary conditions, which restricts the admissible density paths. On $\R^d$, the constructions hold pointwise, but solving the phase equation and identifying the result with the physical evolution require decay, solvability, and uniqueness conditions that we do not pursue.
\end{remark}

The simulations of main-text Figs.~1--3 instead model propagation in unbounded space, where probability that leaves the region of interest never returns. 
If simulated na\"ively on a finite grid, probability could return, e.g., through the opposite face of a periodic grid or by reflection from the grid edge. 
Instead, we surround the detector window $\Odet$ by an absorbing layer and normalize the output over $\Odet$ (Table~\ref{tab:si-boundaries} and Methods). Specifically, the one-dimensional Crank--Nicolson solver builds a perfectly matched layer (PML) into its finite-difference operator, while the two-dimensional split-step Fourier solvers apply the kinetic step in Fourier space, where a PML does not fit, and use a multiplicative absorbing mask instead. Because the layer removes probability, the results for the periodic box do not apply directly. Supplementary \ref{sec:si-absorbing} shows that exact controllability and universal approximation still hold, conditioned on detection in $\Odet$, when the mask is described as an absorbing potential (Proposition~\ref{prop:si-absorbing} and Corollary~\ref{cor:si-detector-window}). We do not analyse the PML separately. Like the mask, it leaves the equation inside $\Odet$ unchanged and removes outgoing probability.

\begin{table}[h]
\caption{Boundary treatment in the analysis and in the simulations. $\Odet$ is the detector window, with the computational domain in parentheses. Lengths are in \si{\micro\meter} for Fig.~3 and dimensionless otherwise.}
\label{tab:si-boundaries}
\small
\begin{ruledtabular}
\begin{tabular}{llll}
 & Solver & Boundary treatment & $\Odet$ (domain)\\
\hline
Theory & exact & periodic; absorbing potential (Note 2.3) & $\Odet\subset\Omega$\\
Figs.~1c, 2 & Crank--Nicolson, 1D & PML, zero field at grid edge & $[-4,4]$ ($[-6,6]$)\\
Fig.~1d & split-step, 2D & periodic grid, absorbing mask & $[-4,4]^2$ ($[-5.5,5.5]^2$)\\
Fig.~3 & split-step, 2D & periodic grid, absorbing mask & $[-40,40]^2$ ($[-55,55]^2$)\\
\end{tabular}
\end{ruledtabular}
\normalsize
\end{table}

\subsection{Madelung equations and potential inversion}
\label{sec:si-madelung}

For a particle of mass $m>0$, a final time $t'>0$, and a real potential $V\in C^\infty(\Omega\times[0,t'])$, the Schr\"odinger equation reads
\begin{equation}
    i\hbar \frac{\partial \psi}{\partial t}
    =
    -\frac{\hbar^2}{2m}\Delta \psi
    + V\psi,
    \qquad (x,t)\in\Omega\times[0,t'] .
    \label{eq:si-schrodinger}
\end{equation}
The dimensionless equation of main-text Figs.~1 and~2 (Methods) corresponds to $\hbar=1$, $m=1/(2A)$ and $V=U_\theta$, and the optical equation of Fig.~3 to $\hbar=1$ and $m=k_0n_{\text{ref}}$, with $z$ in place of $t$ (Supplementary Note~4). The probability density and current are
\begin{equation}
    \rho=\abs{\psi}^2,
    \qquad
    j=\frac{\hbar}{m}\operatorname{Im}\big(\overline{\psi}\,\grad\psi\big),
    \label{eq:si-quantum-current}
\end{equation}
where $\overline{\psi}$ denotes the complex conjugate of $\psi$ and $\operatorname{Im}$ denotes the imaginary part. For every smooth solution, $\partial_t\rho+\Div j=0$.

\begin{lemma}[Conservation and uniqueness]\label{lem:si-uniqueness}
Every solution $\psi\in C^1([0,t'];L^2(\Omega))\cap C([0,t'];H^2(\Omega))$ of equation~\eqref{eq:si-schrodinger} satisfies $\int_\Omega\abs{\psi(x,t)}^2\dd x=\int_\Omega\abs{\psi(x,0)}^2\dd x$, and two such solutions with the same initial state coincide.
\end{lemma}

\begin{proof}
For a solution $\phi$,
$\frac{\dd}{\dd t}\int_\Omega\abs{\phi}^2\dd x
=\frac{2}{\hbar}\operatorname{Im}\int_\Omega\overline{\phi}\big(-\frac{\hbar^2}{2m}\Delta\phi+V\phi\big)\dd x=0$,
because integration by parts on $\Omega$, which has no boundary, gives $\int_\Omega\overline{\phi}\,(-\Delta\phi)\dd x=\int_\Omega\abs{\grad\phi}^2\dd x\in\R$, and $V$ is real. The difference of two solutions with the same initial state is a solution with zero initial norm and therefore vanishes.
\end{proof}

Where $\rho>0$, the wavefunction can be written in Madelung form\cite{madelung1927quantentheorie} $\psi=\sqrt{\rho}\exp(iS/\hbar)$ with a real phase $S$, and then $j=\rho\grad S/m$. Substituting into equation~\eqref{eq:si-schrodinger} and separating real and imaginary parts gives
\begin{subequations}
\begin{align}
    \frac{\partial\rho}{\partial t} + \Div\!\left(\rho\,\frac{\grad S}{m}\right) &= 0,
    \label{eq:si-continuity}\\
    \frac{\partial S}{\partial t} + \frac{1}{2m}\abs{\grad S}^2 + V + Q[\rho] &= 0,
    \label{eq:si-qhj}
\end{align}
\end{subequations}
where
\begin{equation}
    Q[\rho] =
    -\frac{\hbar^2}{2m}\frac{\Delta\sqrt{\rho}}{\sqrt{\rho}}
    \label{eq:si-bohm}
\end{equation}
is the Bohm (quantum) potential. Equation~\eqref{eq:si-continuity} is the continuity equation for the current velocity $\grad S/m=j/\rho$, and equation~\eqref{eq:si-qhj} is the quantum Hamilton--Jacobi equation. Taking the gradient of equation~\eqref{eq:si-qhj} gives
\begin{equation}
    m\left[\frac{\partial v}{\partial t} + (v\cdot\grad)v\right]
    =
    -\grad\left(V+Q[\rho]\right),
    \qquad v=\frac{\grad S}{m},
    \label{eq:si-madelung-velocity}
\end{equation}
so the potential acts as a force on the flow, and $Q[\rho]$ adds a density-dependent quantum pressure. Conversely, a density path and a phase that satisfy the continuity equation determine the potential.

\begin{proposition}[Potential inversion]\label{prop:si-inversion}
Let $\rho\in C^\infty(\Omega\times[0,t'])$ be strictly positive, and let $S\in C^\infty(\Omega\times[0,t'])$ be real and satisfy equation~\eqref{eq:si-continuity}. Define
\begin{equation}
    V
    =
    -\frac{\partial S}{\partial t}
    -\frac{1}{2m}\abs{\grad S}^2
    +\frac{\hbar^2}{2m}\frac{\Delta\sqrt{\rho}}{\sqrt{\rho}} .
    \label{eq:si-potential-inversion}
\end{equation}
Then $V$ is smooth and real, and $\psi=\sqrt{\rho}\exp(iS/\hbar)$ is the unique solution of equation~\eqref{eq:si-schrodinger} with initial state $\sqrt{\rho(\cdot,0)}\exp[iS(\cdot,0)/\hbar]$. In particular, $\abs{\psi(x,t)}^2=\rho(x,t)$ for all $t\in[0,t']$.
\end{proposition}

\begin{proof}
Substitute $\psi$ into equation~\eqref{eq:si-schrodinger} and multiply by $\exp(-iS/\hbar)$. The imaginary part of the residual equals $\hbar/(2\sqrt{\rho})$ times the left-hand side of equation~\eqref{eq:si-continuity}, and the real part equals $-\sqrt{\rho}$ times the left-hand side of equation~\eqref{eq:si-qhj}. The first vanishes by assumption and the second by the definition of $V$. Uniqueness follows from Lemma~\ref{lem:si-uniqueness}.
\end{proof}

Replacing $S$ by $S+\gamma(t)$ (a function of time only, constant in space) multiplies $\psi$ by a global phase and changes $V$ by $-\dd\gamma/\dd t$, but leaves $\rho$ and $j$ unchanged (gauge freedom).

\subsection{Recovering the phase from a density path}
\label{sec:si-phase}

Proposition~\ref{prop:si-inversion} needs a phase for the prescribed density path. Many velocity fields carry the same density path, but the continuity equation~\eqref{eq:si-continuity} determines the one that is a gradient, $\grad S/m$.

\begin{lemma}[Phase recovery]\label{lem:si-phase-recovery}
Let $\rho\in C^\infty(\Omega\times[0,t'])$ be strictly positive, and let $f\in C^\infty(\Omega\times[0,t'])$ satisfy $\int_\Omega f(x,t)\dd x=0$ for all $t$. Then
\begin{equation}
    -\Div\!\left[\rho(x,t)\grad S(x,t)\right]
    =
    f(x,t)
    \label{eq:si-elliptic-phase-lemma}
\end{equation}
has a solution $S\in C^\infty(\Omega\times[0,t'])$, unique up to an additive function of time. If $\int_\Omega\rho(x,t)\dd x=1$ for all $t$, the choice $f=m\,\partial\rho/\partial t$ is admissible, and $\rho$ and the resulting phase satisfy equation~\eqref{eq:si-continuity}.
\end{lemma}

\begin{proof}
By compactness, $\rho\geq c_*>0$. For fixed $t$, the bilinear form $a(g,h)=\int_\Omega\rho\grad g\cdot\grad h\,\dd x$ is bounded on $H^1(\Omega)$ and, by the Poincar\'e inequality, coercive on the subspace of mean-zero functions. By the Lax--Milgram theorem and elliptic regularity, $L_t=-\Div(\rho(\cdot,t)\grad\,\cdot\,)$ therefore maps the mean-zero subspace of $H^{k+2}(\Omega)$ isomorphically onto that of $H^k(\Omega)$ for every $k\geq0$, and $t\mapsto L_t$ is smooth in operator norm. Because inversion is smooth on invertible operators and $f(\cdot,t)$ has zero mean, the mean-zero solution $S(\cdot,t)=L_t^{-1}f(\cdot,t)$ is a smooth function of $t$ with values in every $H^{k+2}(\Omega)$, so $S$ is smooth in $(x,t)$ by Sobolev embedding. The difference of two solutions has vanishing gradient and is therefore a function of time. Finally, $\int_\Omega\rho\,\dd x=1$ implies $\int_\Omega\partial_t\rho\,\dd x=0$.
\end{proof}

\subsection{QMT and flow matching}
\label{sec:si-fm}

Flow matching\cite{Lipman2023} learns a time-dependent velocity field $u(x,t)$ whose flow
\begin{equation}
    \frac{\dd}{\dd t}\phi_t(x)=u(\phi_t(x),t),\qquad \phi_0(x)=x,
    \label{eq:si-flow-map}
\end{equation}
transports samples of a reference distribution at $t=0$ to samples of a target distribution at a final time $t=t'$.
A pair $(\rho,u)$ of a smooth, strictly positive probability density path $\rho(\cdot,t)$ and a smooth velocity field $u(\cdot,t)$ on $\Omega$, $t\in[0,t']$, is a flow-matching path if
\begin{equation}
    \frac{\partial \rho}{\partial t}+\Div(\rho u)=0 .
    \label{eq:si-flow-continuity}
\end{equation}
Along the flow, Liouville's formula $\frac{\dd}{\dd t}\det\grad\phi_t=(\Div u)(\phi_t,t)\det\grad\phi_t$ and equation~\eqref{eq:si-flow-continuity} give $\frac{\dd}{\dd t}\big[\rho(\phi_t(x),t)\det\grad\phi_t(x)\big]=0$, so $\phi_t$ pushes $\rho(\cdot,0)$ forward to $\rho(\cdot,t)$. Comparing equations~\eqref{eq:si-continuity} and~\eqref{eq:si-flow-continuity} gives the correspondence of main-text Fig.~1b: the flow-matching velocity field plays the role of the current velocity $\grad S/m$, and the potential $V$ is the control that generates it through equation~\eqref{eq:si-qhj}.

\begin{proposition}[Nodeless QMT evolutions are flow-matching paths]\label{prop:si-qmt-is-fm}
Let $\psi$ be a smooth, normalized solution of equation~\eqref{eq:si-schrodinger} with $\abs{\psi}>0$ on $\Omega\times[0,t']$, that is, without nodes (zeros of $\psi$). Then $(\rho,v)$ with $\rho=\abs{\psi}^2$ and $v=j/\rho=(\hbar/m)\operatorname{Im}(\grad\psi/\psi)$ is a flow-matching path, and locally $v=\grad S/m$. The flow of $v$, whose trajectories are the de~Broglie--Bohm (pilot-wave) trajectories\cite{bohm1952suggested}, pushes $\abs{\psi(\cdot,0)}^2$ forward to $\abs{\psi(\cdot,t)}^2$.
\end{proposition}

\begin{proof}
Because $j=\rho v$, the conservation law $\partial_t\rho+\Div j=0$ is equation~\eqref{eq:si-flow-continuity} with $u=v$, and the push-forward property follows from Liouville's formula as above.
\end{proof}

\begin{theorem}[Flow-matching density paths are QMT density paths]\label{thm:si-fm-representability}
Let $(\rho,u)$ be a flow-matching path on $\Omega$, and let $S$ solve equation~\eqref{eq:si-elliptic-phase-lemma} with $f=m\,\partial_t\rho$. With $V$ given by equation~\eqref{eq:si-potential-inversion}, $\psi=\sqrt{\rho}\exp(iS/\hbar)$ is the unique solution of equation~\eqref{eq:si-schrodinger} with initial state $\sqrt{\rho(\cdot,0)}\exp[iS(\cdot,0)/\hbar]$, and $\abs{\psi}^2=\rho$ at all times. Its current velocity $v=\grad S/m$ satisfies
\begin{equation}
    \Div\!\left[\rho(v-u)\right]=0 .
    \label{eq:si-hodge}
\end{equation}
If $u=\grad\Phi$ for a smooth function $\Phi$, then $v=u$, so that QMT reproduces the flow-matching velocity field and trajectories, not only the densities.
\end{theorem}

\begin{proof}
By Lemma~\ref{lem:si-phase-recovery}, $\rho$ and $S$ satisfy equation~\eqref{eq:si-continuity}, and Proposition~\ref{prop:si-inversion} gives the first claim. Subtracting equation~\eqref{eq:si-flow-continuity} from equation~\eqref{eq:si-continuity} gives equation~\eqref{eq:si-hodge}. If $u=\grad\Phi$, then $m\Phi$ also solves equation~\eqref{eq:si-elliptic-phase-lemma} with $f=m\,\partial_t\rho$, so $S=m\Phi+\gamma(t)$ by uniqueness, and $v=u$.
\end{proof}

Equation~\eqref{eq:si-hodge} states that $v$ is the $\rho$-weighted Helmholtz projection of $u$ onto gradient fields, $\rho u=\rho v+w$ with $\Div w=0$. A velocity field with nonzero curl cannot be the current velocity of a nodeless wavefunction. For such fields, QMT reproduces the density path but not the trajectories. The path fixes the initial phase $S(\cdot,0)$ up to a constant, but Corollary~\ref{cor:si-real-input} shows that a real input field, as in main-text Figs.~1--3, follows the same densities after a reparameterization of time.

\begin{corollary}[One dimension]\label{cor:si-fm-1d}
Let $(\rho,u)$ be a flow-matching path on $\R$ in place of $\Omega$. Every velocity field on $\R$ is a gradient, $u=\partial\Phi/\partial x$ with $\Phi(x,t)=\int_0^xu(s,t)\dd s$. With $S=m\Phi$, the potential
\begin{equation}
    V=
    -m\frac{\partial\Phi}{\partial t}
    -\frac{m}{2}u^2
    +\frac{\hbar^2}{2m\sqrt{\rho}}
    \frac{\partial^2\sqrt{\rho}}{\partial x^2}
    \label{eq:si-flow-potential}
\end{equation}
reproduces the flow-matching velocity field exactly: $\psi=\sqrt{\rho}\exp(iS/\hbar)$ solves the Schr\"odinger equation on $\R$; identifying it with the physical evolution requires the conditions of Remark~\ref{rem:si-domains}.
\end{corollary}

In higher dimensions, the exact marginal velocity fields of the most widely used flow-matching and diffusion models are also gradients.

\begin{proposition}[Gaussian probability paths]\label{prop:si-gaussian-paths}
Let $\pi$ be a probability measure on $\R^d$ with finite moments of all orders, let $\alpha,\sigma\in C^\infty([0,1])$ with $\sigma>0$, and let $p_t$ be the density of $X_t=\alpha_tX_1+\sigma_tX_0$, where $X_1\sim\pi$ and $X_0\sim\mathcal N(0,I_d)$ are independent. 
Then the marginal flow-matching velocity field
\begin{equation}
    u_t(x)=\E\!\left[\dot\alpha_tX_1+\dot\sigma_tX_0\,\middle|\,X_t=x\right]
\end{equation}
is a gradient field, $u_t=\grad\Phi_t$, with
\begin{subequations}
\label{eq:si-gaussian-path-potential}
\begin{align}
    \Phi_t(x)&=\frac{\dot\sigma_t}{2\sigma_t}\abs{x}^2
    +\left(\dot\alpha_t-\frac{\alpha_t\dot\sigma_t}{\sigma_t}\right)\Psi_t(x),\\
    \Psi_t(x)&=\frac{\sigma_t^2}{\alpha_t}\log\int_{\R^d}
    \exp\!\left(\frac{\alpha_t\langle x,y\rangle-\alpha_t^2\abs{y}^2/2}{\sigma_t^2}\right)\pi(\dd y),
\end{align}
\end{subequations}
where $\Psi_t(x)=\langle\E X_1,x\rangle$ if $\alpha_t=0$.
\end{proposition}

\begin{proof}
Since $X_0=(X_t-\alpha_tX_1)/\sigma_t$, we have $u_t(x)=(\dot\alpha_t-\alpha_t\dot\sigma_t/\sigma_t)\,\E[X_1\mid X_t=x]+(\dot\sigma_t/\sigma_t)\,x$. For $\alpha_t\neq0$, Bayes' rule gives the conditional law of $X_1$ given $X_t=x$: because $X_t$ given $X_1=y$ is $\mathcal N(\alpha_ty,\sigma_t^2I_d)$, it is proportional to 
$\exp[(\alpha_t\langle x,y\rangle-\alpha_t^2\abs{y}^2/2)/\sigma_t^2]\,\pi(\dd y)$ after dropping the $y$-independent factor $\exp(-\abs{x}^2/2\sigma_t^2)$, and differentiation under the integral gives $\E[X_1\mid X_t=x]=\grad\Psi_t(x)$ (Tweedie's formula\cite{efron2011tweedie}). For $\alpha_t=0$, $X_1$ and $X_t$ are independent and $\E[X_1\mid X_t=x]=\E X_1$. Replacing $\alpha_t$ by a real variable $a$ turns the integral into a positive function $Z(x,a,t)$ with $Z(x,0,t)=1$. Because $a\langle x,y\rangle-a^2\abs{y}^2/2\leq\abs{x}^2/2$, every derivative of its integrand in $(x,a,t)$ is bounded by $\exp(\abs{x}^2/2\sigma_t^2)$ times a polynomial in $(x,y,a)$, so finite moments of $\pi$ justify the differentiation and make $Z$ smooth. By Hadamard's lemma, $\log Z=aG$ with $G$ smooth and $\sigma_t^2G(x,0,t)=\langle\E X_1,x\rangle$, so $\Psi_t=\sigma_t^2G(\cdot,\alpha_t,t)$ is smooth in $(x,t)$, also where $\alpha_t=0$, and so is $p_t=(2\pi\sigma_t^2)^{-d/2}\exp(-\abs{x}^2/2\sigma_t^2)\,Z(\cdot,\alpha_t,t)$.
\end{proof}

The usual endpoint conditions $\alpha_0=0$, $\sigma_0=1$, $\alpha_1=1$ and $\sigma_1=\sigma_{\min}\geq0$ make $p_0=\mathcal N(0,I_d)$ and $p_1\approx\pi$; the gradient property does not require them. Proposition~\ref{prop:si-gaussian-paths} covers the Gaussian conditional paths of flow matching\cite{Lipman2023} ($\alpha_t=t$, $\sigma_t=1-(1-\sigma_{\min})t$ with $\sigma_{\min}>0$) and, for $t<1$, rectified flow\cite{Liu2022} ($\sigma_{\min}=0$). For diffusion models $\dd Y_t=b(Y_t,t)\dd t+g(t)\dd B_t$ with linear drift, $b(x,t)=-\tfrac12\beta(t)x$ or $b=0$ (ref.~\citenum{Song2021}), the probability-flow velocity $b(x,t)-\tfrac12g(t)^2\grad\log p_t(x)$, with $p_t$ the density of $Y_t$, is also a gradient field. For all these models, $(p_t,u_t)$ satisfies the continuity equation~\eqref{eq:si-flow-continuity}, and QMT therefore reproduces the velocity field and trajectories in any dimension, although a trained velocity network is only approximately a gradient field. Because $p_t>0$, the wavefunction $\psi_t=\sqrt{p_t}\exp(im\Phi_t/\hbar)$ solves the Schr\"odinger equation on $\R^d$ for $t'=1$ with the smooth potential of equation~\eqref{eq:si-potential-inversion}. For general $t'$, replace $p_t$ and $\Phi_t$ by $p_{t/t'}$ and $\Phi_{t/t'}/t'$. Identifying this solution with the physical evolution requires decay and uniqueness conditions that we do not verify (Remark~\ref{rem:si-domains}). The required initial phase is generally not flat: for $\alpha_0=0$ and $\sigma_0=1$, $S(x,0)=(m/t')\big[\tfrac12\dot\sigma_0\abs{x}^2+\dot\alpha_0\langle\E X_1,x\rangle\big]$, a lens-like quadratic phase with a linear tilt.

\begin{remark}[Trained potentials]\label{rem:si-trained}
The potentials in Figs.~1--3 in the main text are trained on the terminal density only (Methods), so they select one of the many density paths that reach the target, not necessarily the path of a standard flow-matching model or the optimal-transport path.
\end{remark}

\section{Controllability, universal approximation, and optimal transport}
\label{sec:si-controllability}

This note proves that, from any smooth, strictly positive input density, QMT reaches smooth, strictly positive targets exactly (\ref{sec:si-exact}) and approximates all other targets, including empirical distributions, to arbitrary accuracy in $W_2$ (Theorem~\ref{thm:si-universal-approx} in \ref{sec:si-universal}). \ref{sec:si-absorbing} extends both results to the output conditioned on detection in a window with an absorbing mask, and \ref{sec:si-ot} describes the optimal-transport path.

\subsection{Exact controllability between smooth positive densities}
\label{sec:si-exact}

The input field $\sqrt{\rho_0}\exp(iS_0/\hbar)$ fixes the initial slope $\partial_t\rho(\cdot,0)=-\Div(\rho_0\grad S_0/m)$ of the density path, so we first construct positive paths with a prescribed initial slope.

\begin{proposition}[Positive interpolants with a prescribed initial slope]
\label{prop:si-compatible-path}
Let $\rho_0,\rho_{t'}\in C^\infty(\Omega)$ be strictly positive probability densities, and let $g_0\in C^\infty(\Omega)$ satisfy $\int_\Omega g_0\dd x=0$. There is a strictly positive probability density path $\rho\in C^\infty(\Omega\times[0,t'])$ with
\[
    \rho(\cdot,0)=\rho_0,\qquad
    \rho(\cdot,t')=\rho_{t'},\qquad
    \frac{\partial \rho}{\partial t}(\cdot,0)=g_0 .
\]
\end{proposition}

\begin{proof}
Write $s=t/t'$ and let
\begin{equation}
    \tilde\rho(x,t)
    =
    \big[(1-s^2)\rho_0(x)+s^2\rho_{t'}(x)\big]\exp\!\big[t(1-s)\,g_0(x)/\rho_0(x)\big].
\end{equation}
Then $\tilde\rho$ is smooth and strictly positive, $\tilde\rho(\cdot,0)=\rho_0$, $\tilde\rho(\cdot,t')=\rho_{t'}$ and $\partial_t\tilde\rho(\cdot,0)=g_0$. The path $\rho=\tilde\rho/\int_\Omega\tilde\rho\dd x$ has the required properties, because the normalization equals one at $t=0$ and $t=t'$ and its derivative at $t=0$ is $\int_\Omega g_0\dd x=0$.
\end{proof}

\begin{theorem}[Density controllability]\label{thm:si-density-controllability}
Let $\rho_0,\rho_{t'}\in C^\infty(\Omega)$ be strictly positive probability densities, let $S_0\in C^\infty(\Omega)$, and let $\rho\in C^\infty(\Omega\times[0,t'])$ be a strictly positive probability density path with $\rho(\cdot,0)=\rho_0$, $\rho(\cdot,t')=\rho_{t'}$ and
\begin{equation}
    \frac{\partial \rho}{\partial t}(\cdot,0)=-\Div\!\left(\rho_0\frac{\grad S_0}{m}\right).
    \label{eq:si-initial-compatibility}
\end{equation}
Let $S$ solve equation~\eqref{eq:si-elliptic-phase-lemma} with $f=m\,\partial_t\rho$, shifted by a constant so that $S(\cdot,0)=S_0$, and define $V$ by equation~\eqref{eq:si-potential-inversion}. Then the unique solution of equation~\eqref{eq:si-schrodinger} with $\psi(\cdot,0)=\sqrt{\rho_0}\exp(iS_0/\hbar)$ satisfies $\abs{\psi(\cdot,t)}^2=\rho(\cdot,t)$ for all $t\in[0,t']$, and in particular $\abs{\psi(\cdot,t')}^2=\rho_{t'}$. By Proposition~\ref{prop:si-compatible-path} with $g_0=-\Div(\rho_0\grad S_0/m)$, which has zero mean, such a path exists for every pair $\rho_0,\rho_{t'}$ and every $S_0$.
\end{theorem}

\begin{proof}
By equation~\eqref{eq:si-initial-compatibility}, $S_0$ solves equation~\eqref{eq:si-elliptic-phase-lemma} at $t=0$. By the uniqueness part of Lemma~\ref{lem:si-phase-recovery}, $S(\cdot,0)$ and $S_0$ differ by a constant, and adding this constant to $S$ for all $t$ preserves equation~\eqref{eq:si-elliptic-phase-lemma}. Proposition~\ref{prop:si-inversion} completes the proof.
\end{proof}

For a real input field, condition~\eqref{eq:si-initial-compatibility} requires $\partial_t\rho(\cdot,0)=0$, which a reparameterization of time can impose on any density path, as the following corollary shows.

\begin{corollary}[Real input fields and time reparameterization]\label{cor:si-real-input}
Let $S_0$ be constant, as for the real Gaussian input fields of main-text Figs.~1--3. Let $\widehat\rho\in C^\infty(\Omega\times[0,1])$ be any strictly positive probability density path from $\rho_0$ to $\rho_{t'}$, and let $\widehat S$ solve equation~\eqref{eq:si-elliptic-phase-lemma} for $\widehat\rho$ with $f=m\,\partial_s\widehat\rho$, where $s\in[0,1]$ is the path parameter (Lemma~\ref{lem:si-phase-recovery}). Let $\tau\in C^\infty([0,t'])$ take values in $[0,1]$ and satisfy $\tau(0)=0$, $\tau(t')=1$ and $\tau'(0)=0$, for example the nondecreasing $\tau(t)=3(t/t')^2-2(t/t')^3$. Then the density path $\rho(x,t)=\widehat\rho(x,\tau(t))$ is realized from $\sqrt{\rho_0}\exp(iS_0/\hbar)$ by the potential
\begin{equation}
    V(x,t)=
    -\tau''(t)\,\widehat S
    -\tau'(t)^2\left[\frac{\partial\widehat S}{\partial s}+\frac{1}{2m}\abs{\grad\widehat S}^2\right]
    -Q\big[\widehat\rho(\cdot,\tau(t))\big],
    \label{eq:si-reparameterized-potential}
\end{equation}
where $\widehat S$ and its derivatives are evaluated at $(x,\tau(t))$.
\end{corollary}

\begin{proof}
By the chain rule, $S(x,t)=S_0+\tau'(t)\widehat S(x,\tau(t))$ satisfies equation~\eqref{eq:si-continuity} for $\rho$, and $S(\cdot,0)=S_0$ because $\tau'(0)=0$. Proposition~\ref{prop:si-inversion}, with $V$ computed from equation~\eqref{eq:si-potential-inversion}, gives equation~\eqref{eq:si-reparameterized-potential}.
\end{proof}

For nondecreasing $\tau$, only the speed along the path changes, not the sequence of intermediate densities or the terminal density. QMT can therefore follow every smooth, strictly positive density path from $\rho_0$, including optimal-transport and flow-matching paths, from a real input field, up to this reparameterization of time.

\subsection{Universal approximation}
\label{sec:si-universal}

For targets that are not smooth and strictly positive, such as empirical measures and densities with zeros, we measure the approximation error with the quadratic Wasserstein distance
\begin{equation}
    W_2^2(\mu,\nu)
    =
    \inf_{\gamma\in\Pi(\mu,\nu)}
    \int_{\Omega\times\Omega}d_\Omega(x,y)^2\,\dd\gamma(x,y),
    \label{eq:si-w2-definition}
\end{equation}
where $\Pi(\mu,\nu)$ is the set of couplings of $\mu$ and $\nu$ and $d_\Omega$ is the periodic distance on $\Omega$. On the compact space $\Omega$, convergence in $W_2$ is equivalent to weak convergence of probability measures (see Theorem~6.9 in ref.~\citenum{villani2009optimal}).

\begin{proposition}[Smooth positive approximation of arbitrary targets]
\label{prop:si-smooth-approx}
Let $\mu$ be any probability measure on $\Omega$. For every $\delta>0$ there is a strictly positive probability density $\rho^\delta\in C^\infty(\Omega)$ with $W_2(\rho^\delta\dd x,\mu)<\delta$.
\end{proposition}

\begin{proof}
Let $K_\sigma(x,y)=\sum_{k\in\mathbb{Z}^d}(4\pi\sigma)^{-d/2}\exp\!\big(-\abs{x-y-kL}^2/(4\sigma)\big)$ be the heat kernel on $\mathbb{T}^d_L$ at time $\sigma>0$, and set
\begin{equation}
    \rho^\sigma(x)=\int_\Omega K_\sigma(x,y)\,\dd\mu(y).
    \label{eq:si-smooth-positive-approx}
\end{equation}
The heat kernel is smooth, strictly positive and satisfies $\int_\Omega K_\sigma(x,y)\dd x=1$, so $\rho^\sigma$ is a smooth, strictly positive probability density. Because $K_\sigma(\cdot,y)$ is the law of $y+\sqrt{2\sigma}Z$ (mod $L$) with $Z\sim\mathcal N(0,I_d)$, coupling $Y\sim\mu$ with $Y+\sqrt{2\sigma}Z$, where $Z$ is independent of $Y$, gives $W_2(\rho^\sigma\dd x,\mu)\leq\sqrt{2d\sigma}$, which is less than $\delta$ for $\sigma<\delta^2/(2d)$.
\end{proof}

\begin{theorem}[Universal density approximation by QMT]
\label{thm:si-universal-approx}
Let $\rho_0\in C^\infty(\Omega)$ be a strictly positive probability density, $S_0\in C^\infty(\Omega)$ an initial phase, and $\mu$ any probability measure on $\Omega$. For every $\delta>0$ there is a smooth real potential $V_\delta$ such that the solution of equation~\eqref{eq:si-schrodinger} with $\psi(\cdot,0)=\sqrt{\rho_0}\exp(iS_0/\hbar)$ has a terminal density $\rho^\delta_{t'}$ with
\[
    W_2(\rho_{t'}^\delta\dd x,\mu)<\delta .
\]
\end{theorem}

\begin{proof}
Choose $\rho_{t'}^\delta$ by Proposition~\ref{prop:si-smooth-approx} and apply Theorem~\ref{thm:si-density-controllability}.
\end{proof}

The theorem does not bound $V_\delta$ uniformly as $\delta\to0$. Sharper targets generally require stronger or faster-varying potentials.

\subsection{Absorbing layers and detector windows}
\label{sec:si-absorbing}

For the simulations in Figs.~1--3 of the main text, an absorbing layer surrounds the detector window $\Odet$ (Supplementary \ref{sec:si-setting}). We model absorption by a complex absorbing potential $-iW$ with $W\in C^\infty(\Omega)$ and $W\geq0$ (in the simulations, $W=0$ on $\Odet$),
\begin{equation}
    i\hbar \frac{\partial \psi}{\partial t}
    =
    -\frac{\hbar^2}{2m}\Delta \psi
    + \left(V-iW\right)\psi .
    \label{eq:si-schrodinger-absorbing}
\end{equation}
In Madelung variables, the quantum Hamilton--Jacobi equation~\eqref{eq:si-qhj} is unchanged, whereas the continuity equation acquires a sink,
\begin{equation}
    \frac{\partial\rho}{\partial t}+\Div\!\left(\rho\,\frac{\grad S}{m}\right)=-\frac{2}{\hbar}W\rho .
    \label{eq:si-continuity-absorbing}
\end{equation}
The uniqueness part of Lemma~\ref{lem:si-uniqueness} holds for equation~\eqref{eq:si-schrodinger-absorbing}, because now $\frac{\dd}{\dd t}\int_\Omega\abs{\phi}^2\dd x=-\frac2\hbar\int_\Omega W\abs{\phi}^2\dd x\leq0$.

\begin{proposition}[Controllability with absorption]\label{prop:si-absorbing}
Let $W$ be as above, let $\rho_0,\rho_{t'}\in C^\infty(\Omega)$ be strictly positive probability densities, and let $S_0\in C^\infty(\Omega)$. There is a smooth real potential $V$ such that the solution of equation~\eqref{eq:si-schrodinger-absorbing} with $\psi(\cdot,0)=\sqrt{\rho_0}\exp(iS_0/\hbar)$ satisfies $\abs{\psi(\cdot,t')}^2=M\rho_{t'}$, where $M\in(0,1]$ is the probability of not being absorbed. The terminal density conditioned on survival is therefore exactly $\rho_{t'}$, and conditioned on detection in $\Odet$ it is $\rho_{t'}/\int_{\Odet}\rho_{t'}\dd x$ on $\Odet$.
\end{proposition}

\begin{proof}
Let $\bar\rho$ be the path of Proposition~\ref{prop:si-compatible-path} with $g_0=-\Div(\rho_0\grad S_0/m)-\frac2\hbar\big(W-\langle W\rangle_0\big)\rho_0$, where $\langle W\rangle_0=\int_\Omega W\rho_0\dd x$, so that $\int_\Omega g_0\dd x=0$. Set $\rho=M(t)\bar\rho$ with \linebreak $M(t)=\exp\big[-\frac2\hbar\int_0^t\int_\Omega W\bar\rho\dd x\dd s\big]$. Then $\frac{\dd}{\dd t}\int_\Omega\rho\dd x=-\frac2\hbar\int_\Omega W\rho\dd x$, so $f=m(\partial_t\rho+\frac2\hbar W\rho)$ has zero mean, and Lemma~\ref{lem:si-phase-recovery} gives a smooth phase $S$ for which $\rho$ and $S$ satisfy equation~\eqref{eq:si-continuity-absorbing}. At $t=0$, the choice of $g_0$ gives $f(\cdot,0)=-\Div(\rho_0\grad S_0)$, so $S$ can be shifted by a constant to satisfy $S(\cdot,0)=S_0$. With $V$ from equation~\eqref{eq:si-potential-inversion}, which is unchanged by $W$, $\psi=\sqrt{\rho}\exp(iS/\hbar)$ solves equation~\eqref{eq:si-schrodinger-absorbing}: the real part of the residual is as in Proposition~\ref{prop:si-inversion}, and the imaginary part is equation~\eqref{eq:si-continuity-absorbing}. Uniqueness identifies $\psi$ with the evolution of the initial state, and $M=M(t')$.
\end{proof}

\begin{corollary}[Detector windows]\label{cor:si-detector-window}
Let $\Odet\subset\Omega$ be closed and convex with nonempty interior, such as the centred windows of Table~\ref{tab:si-boundaries}, let $W$, $\rho_0$ and $S_0$ be as in Proposition~\ref{prop:si-absorbing}, and let $\mu$ be a probability measure on $\Omega$ with $\mu(\Odet)=1$. For every $\delta>0$ there is a smooth real potential for which the terminal density of equation~\eqref{eq:si-schrodinger-absorbing}, conditioned on detection in $\Odet$, lies within $\delta$ of $\mu$ in $W_2$ with respect to the Euclidean distance on $\Odet$.
\end{corollary}

\begin{proof}
Contracting $\mu$ towards an interior point of $\Odet$ by a factor $1-\epsilon$ moves no mass by more than $\epsilon\operatorname{diam}\Odet$ and, because $\Odet$ is convex, places all mass in the interior, so we may assume $\mu(\partial\Odet)=0$. The densities $\rho^\sigma_{t'}$ of equation~\eqref{eq:si-smooth-positive-approx} converge weakly to $\mu$. Because $\mu(\partial\Odet)=0$ and $\mu(\Odet)=1$, their normalized restrictions to $\Odet$ also converge weakly to $\mu$ (portmanteau theorem), hence in $W_2$ on the compact set $\Odet$. Proposition~\ref{prop:si-absorbing} realizes each $\rho^\sigma_{t'}$.
\end{proof}

The corollary thus covers empirical measures of samples clipped to $\partial\Odet$ (Methods).

The split-step solver of main-text Fig.~1d applies the mask $M_{\mathrm{abs}}=\exp(-\gamma_{\mathrm{abs}}d_{\mathrm{det}}^2)$ after every step $\Delta t$ (Methods), where $d_{\mathrm{det}}(x)$ is the Euclidean distance from $x$ to $\Odet$. This is a first-order splitting of equation~\eqref{eq:si-schrodinger-absorbing} with $W=\hbar\gamma_{\mathrm{abs}}d_{\mathrm{det}}^2/\Delta t$. This $W$ is continuous but not smooth: $d_{\mathrm{det}}^2$ is only $C^{1,1}$ across $\partial\Odet$ and has a kink at the edge of the periodic box. Replacing $W$ by a smooth $W_\epsilon\geq0$ with $\sup\abs{W_\epsilon-W}\leq\epsilon$ changes the solution by at most $\epsilon t'/\hbar$ in $L^2(\Omega)$, for every real potential and normalized initial state, because both evolutions are $L^2$ contractions (Duhamel's formula). For a fixed $\rho_{t'}$, the construction of Proposition~\ref{prop:si-absorbing} with $W_\epsilon$ gives a detection probability of at least $e^{-2t'(\sup W+\epsilon)/\hbar}\int_{\Odet}\rho_{t'}\dd x$, bounded away from zero as $\epsilon\to0$, so Corollary~\ref{cor:si-detector-window} also holds for this non-smooth $W$. The PML of the one-dimensional solver (main-text Figs.~1c and~2) is not covered by this analysis (Supplementary \ref{sec:si-setting}).

\subsection{The optimal-transport path}
\label{sec:si-ot}

Theorem~\ref{thm:si-density-controllability} allows any density path from $\rho_0$ to $\rho_{t'}$. Quadratic optimal transport (OT) singles out a canonical one, along which the QMT potential reduces to the negative Bohm potential of the intermediate densities and its strength can be bounded explicitly (equation~\eqref{eq:si-ot-potential} and Remark~\ref{rem:si-amplitude}). The OT path minimizes the kinetic action\cite{benamou2000computational}
\begin{equation}
    \mathcal{A}[\rho,v]
    =
    \int_0^{t'}\int_\Omega \frac{1}{2}m\,\rho(x,t)\abs{v(x,t)}^2\dd x\,\dd t
    \label{eq:si-bb-action}
\end{equation}
subject to
\begin{equation}
    \frac{\partial \rho}{\partial t}+\Div(\rho v)=0,\qquad
    \rho(\cdot,0)=\rho_0,\quad \rho(\cdot,t')=\rho_{t'} .
\end{equation}

\begin{proposition}[Smooth OT interpolation on the flat torus]\label{prop:si-ot-torus}
Let $\rho_0,\rho_{t'}\in C^\infty(\mathbb{T}^d_L)$ be strictly positive probability densities. The OT interpolation $\rho_{\mathrm{OT}}$ between them is a smooth, strictly positive probability density path, and its velocity is $v_{\mathrm{OT}}=\grad S_{\mathrm{OT}}/m$ with a smooth, single-valued phase that satisfies
\begin{equation}
    \frac{\partial S_{\mathrm{OT}}}{\partial t}
    +\frac{1}{2m}\abs{\grad S_{\mathrm{OT}}}^2=0 .
    \label{eq:si-ot-hj}
\end{equation}
\end{proposition}

\begin{proof}
By the theory of OT on the flat torus\cite{cordero1999transport}, which combines Brenier's theorem\cite{brenier1991polar} with Caffarelli's regularity theory\cite{caffarelli1992regularity}, the OT map is $T(x)=x-\grad\varphi(x)$ with a smooth, periodic $\varphi$ for which $\abs{x}^2/2-\varphi$ is convex. The Monge--Amp\`ere equation $\det(I-\grad^2\varphi)=\rho_0/(\rho_{t'}\circ T)>0$ then shows that $I-\grad^2\varphi$ is positive definite. For $s\in[0,1]$, the maps $T_s(x)=x-s\grad\varphi(x)$ commute with lattice translations, and their Jacobians $I-s\grad^2\varphi=(1-s)I+s(I-\grad^2\varphi)$ are positive definite, uniformly in $x$ by periodicity. Thus $T_s$, the gradient of the uniformly convex function $\abs{x}^2/2-s\varphi$, is a diffeomorphism of $\R^d$ and therefore of the torus, and $\rho_{\mathrm{OT}}(\cdot,st')=(T_s)_\#\rho_0$ is smooth and strictly positive. The function $\Phi_s$ defined by \linebreak $\Phi_s(T_s(x))=-\varphi(x)+\frac s2\abs{\grad\varphi(x)}^2$ is smooth and periodic, satisfies $\grad\Phi_s(T_s(x))=-\grad\varphi(x)$, which is the velocity of $T_s(x)$, and solves $\partial_s\Phi_s+\frac12\abs{\grad\Phi_s}^2=0$. Setting $S_{\mathrm{OT}}(x,t)=(m/t')\,\Phi_{t/t'}(x)$ gives the claim.
\end{proof}

With the phase of equation~\eqref{eq:si-ot-hj}, the kinetic and time-derivative terms cancel in equation~\eqref{eq:si-potential-inversion}, and the QMT potential that follows the OT path is the negative Bohm potential,
\begin{equation}
    V_{\mathrm{OT}}(x,t)
    =
    \frac{\hbar^2}{2m}
    \frac{\Delta\sqrt{\rho_{\mathrm{OT}}}}{\sqrt{\rho_{\mathrm{OT}}}}
    =-Q[\rho_{\mathrm{OT}}] .
    \label{eq:si-ot-potential}
\end{equation}
This realization requires the initial phase $S_{\mathrm{OT}}(\cdot,0)=-(m/t')\varphi$, which is not constant unless $\rho_0=\rho_{t'}$. From a real input field, Corollary~\ref{cor:si-real-input} applied to the OT path parameterized by $s=t/t'\in[0,1]$, with phase $\widehat S=m\Phi_s$, gives
\begin{equation}
    V(x,t)=-Q\big[\rho_{\mathrm{OT}}(\cdot,\tau(t)t')\big](x)-\tau''(t)\,\widehat S(x,\tau(t)).
    \label{eq:si-ot-potential-real}
\end{equation}

\begin{remark}[Potential amplitude and propagation length]\label{rem:si-amplitude}
Up to a spatially constant gauge term, equation~\eqref{eq:si-ot-potential-real} gives
$\sup_x\abs{V(x,t)}\leq\sup_x\abs{Q[\rho_{\mathrm{OT}}(\cdot,\tau t')]}+\tfrac12\abs{\tau''(t)}\operatorname{osc}_x\widehat S(\cdot,\tau)$, where $\operatorname{osc}$ denotes the difference between maximum and minimum. For $\tau(t)=\tau_1(t/t')$ with a fixed profile $\tau_1$ on $[0,1]$, such as $\tau_1(s)=3s^2-2s^3$, the first term does not depend on $t'$, whereas the second decreases as $1/t'^2$. The first term describes the quantum pressure of the intermediate densities, and the second describes the force needed to set the flow in motion from rest. Longer propagation therefore lowers this bound on the potential, down to the level set by the quantum pressure. OT minimizes the kinetic action, not a control cost such as $\sup\abs{V}$, so other paths to the same target may require weaker potentials.
\end{remark}

\begin{remark}[Schr\"odinger bridges and diffusion models]\label{rem:si-sb}
The Schr\"odinger bridge between $\rho_0$ and $\rho_{t'}$ is defined as the stochastic process that is closest in relative entropy to Brownian motion with diffusivity $\nu$ among all processes with these endpoint marginals\cite{leonard2014survey}. Optimal transport is its zero-noise limit\cite{leonard2012schrodinger}, and Schr\"odinger bridges underlie several diffusion-based generative models\cite{debortoli2021diffusion}. When the Schr\"odinger factors of the bridge are smooth and positive, its marginals are $\rho=\zeta\hat\zeta$, with $\partial_t\hat\zeta=\nu\Delta\hat\zeta$ and $\partial_t\zeta=-\nu\Delta\zeta$. These marginals satisfy the continuity equation with the gradient velocity $v=\nu\grad\log(\zeta/\hat\zeta)$, because \linebreak $\partial_t(\zeta\hat\zeta)=\nu(\zeta\Delta\hat\zeta-\hat\zeta\Delta\zeta)=-\Div(\rho v)$. By Theorem~\ref{thm:si-fm-representability}, QMT therefore reproduces the marginals and the velocity field of the bridge, starting from the phase $m\nu\log(\zeta/\hat\zeta)$ at $t=0$. The random sample paths of the bridge, however, differ from the de~Broglie--Bohm trajectories.
\end{remark}

\section{Triangular transport and conditional sampling}
\label{sec:si-triangular-transport}

The main text uses triangular transport in two ways. In the simulations of Fig.~2, each factor $\rho(x_k\mid x_1,\ldots,x_{k-1})$ of the telescoping factorization of the target density (main-text equation~(2)) is approximated by the terminal density of a one-dimensional QMT evolution whose potential depends on the previously measured coordinates (\ref{sec:si-kr-qmt}). In this case, the triangular map is never written down. The quantum evolutions implement the inverses of its components (Proposition~\ref{prop:si-monotone}). In the experiment of Fig.~4, a triangular map is fitted digitally in the 16-dimensional latent space of an autoencoder, and the conditional densities of successive coordinate pairs are displayed on a spatial light modulator (SLM) and sampled with single photons (\ref{sec:si-mnist}).

\subsection{Triangular maps and sequential sampling}
\label{sec:si-kr}

Let $\pi=\rho_{\mathrm{tar}}$ and $\eta$ be strictly positive target and reference densities on $\R^K$, where $K$ is the number of coordinates ($D$ in the main text). The bold $\bS$ denotes a transport map, an entirely different object than the phase $S$ of Supplementary Notes~1 and~2. We use the superscript $\sharp$ to denote the pullback of a density
and the subscript $\#$ to denote the pushforward of a density. For a monotone triangular map $\bS:\R^K\to\R^K$ (equation~\eqref{eq:si-triangular-map}), whose components are continuously differentiable bijections of $\R$ in their last argument with positive derivatives, the model density is the pullback
\begin{equation}
    \rho_{\bS}(\bx)=\bS^\sharp\eta(\bx)
    =\eta(\bS(\bx))\abs{\det\nabla_{\bx}\bS(\bx)}.
    \label{eq:si-pullback}
\end{equation}
If $\bS_\#\pi=\eta$, then $\rho_{\bS}=\pi$. The triangular structure is
\begin{equation}
    \bS(\bx)=
    \begin{bmatrix*}[l]
    S_1(x_1)\\
    S_2(x_1,x_2)\\
    \qquad \vdots\\
    S_K(x_1,\ldots,x_K)
    \end{bmatrix*},
    \qquad
    \det\nabla_{\bx}\bS=\prod_{k=1}^{K}\frac{\partial S_k}{\partial x_k},
    \qquad \frac{\partial S_k(x_{1:k})}{\partial x_k}>0,
    \label{eq:si-triangular-map}
\end{equation}
where $x_{1:k}=(x_1,\ldots,x_k)$. With $S_k^{-1}(\cdot\,;x_{1:k-1})$ the inverse of $S_k$ in its last argument, $\bS$ is inverted sequentially, and drawing $\bz\sim\eta$ gives a sample of $\rho_{\bS}$:
\begin{align}
    x_1 &= S_1^{-1}(z_1),\nonumber\\
    x_2 &= S_2^{-1}(z_2;x_1),\nonumber\\
    &\ \vdots\nonumber\\
    x_K &= S_K^{-1}(z_K;x_1,\ldots,x_{K-1}).
    \label{eq:si-recursive-inverse}
\end{align}
For the product reference $\eta(\bz)=\prod_{k=1}^K\eta_1(z_k)$ with the standard normal density $\eta_1$, the unique monotone triangular map that pushes $\pi$ forward to $\eta$ is the Knothe--Rosenblatt rearrangement\cite{Santambrogio2015,Ramgraber2025}
\begin{equation}
    S_k(x_{1:k})=
    \Phi^{-1}\!\left(F_k(x_k\mid x_{1:k-1})\right),
    \label{eq:si-kr-cdf-map}
\end{equation}
where $F_k$ is the conditional cumulative distribution function (CDF) of the $k$-th marginal of the target (simply the marginal CDF of $x_1$ for $k=1$) and $\Phi$ is the standard normal CDF. The map depends on the ordering of the coordinates. If these CDFs are continuously differentiable with positive derivatives in $x_k$, so is $S_k$, a bijection of $\R$ in $x_k$, and $\rho_{\bS}=\pi$. For any monotone triangular map and this product reference, the model conditionals are
\begin{equation}
    \rho_{\bS}(x_k\mid x_{1:k-1})=
    \eta_1\!\left(S_k(x_{1:k})\right)\frac{\partial S_k(x_{1:k})}{\partial x_k},
    \label{eq:si-map-conditional}
\end{equation}
because $\rho_{\bS}$ is the product of these factors (equations~\eqref{eq:si-pullback} and~\eqref{eq:si-triangular-map}) and each factor is normalized in $x_k$, as the substitution $z_k=S_k(x_{1:k})$ shows: integrating out the later coordinates successively leaves the product of the first $k$ factors as the marginal density of $x_{1:k}$. Hence
\begin{equation}
    \rho_{\bS}(\bx)=\rho_{\bS}(x_1)
    \prod_{k=2}^{K}\rho_{\bS}(x_k\mid x_{1:k-1}),
    \label{eq:si-chain-rule}
\end{equation}
which is main-text equation~(2) for the model density.

\subsection{Learning triangular maps from samples}
\label{sec:si-kr-fit}

A monotone parameterization is
\begin{equation}
    S_k(x_{1:k})=g_k(x_{1:k-1})+
    \int_0^{x_k}r_k\!\left(h_k(x_{1:k-1},s)\right)\dd s,
    \label{eq:si-monotone-param}
\end{equation}
where $g_k$ and $h_k$ are continuous (for example, ReLU networks) and $r_k$ is a smooth positive function\cite{Baptista2024}. Then $\partial S_k/\partial x_k=r_k(h_k)>0$ exists and is continuous. A uniform positive lower bound on $r_k$ implies $S_k(x_{1:k-1},s)\to\pm\infty$ as $s\to\pm\infty$, so each component maps onto the whole real line. Given samples from $\pi$, the map is estimated by minimizing
\begin{equation}
    D_{\mathrm{KL}}\!\left(\pi\,\Vert\,\rho_{\bS}\right)
    =\E_\pi\!\left[
    \log\frac{\pi(\bx)}{\eta(\bS(\bx))\abs{\det\nabla\bS(\bx)}}
    \right].
\end{equation}
If the target entropy is finite, dropping it gives the negative log-likelihood
\begin{equation}
    \bS_{\mathrm{opt}}=
    \argmin_{\bS\in\mathcal F}\E_\pi\!\left[
    -\log\eta(\bS(\bx))-\log\abs{\det\nabla\bS(\bx)}\right],
    \label{eq:si-triangular-kl}
\end{equation}
where $\mathcal F$ is an admissible class of triangular maps. For the standard normal reference and samples $\{\bx^i\}_{i=1}^N$, the empirical objective is, up to an additive constant,
\begin{equation}
    \mathcal J(\bS)=\frac1N\sum_{i=1}^N\sum_{k=1}^K
    \left[\frac12S_k(x_{1:k}^i)^2-
    \log\frac{\partial S_k(x_{1:k}^i)}{\partial x_k}\right].
    \label{eq:si-triangular-sample-loss}
\end{equation}
The QMT potentials of main-text Fig.~2 are trained with the same negative log-likelihood, plus a regularization term (Methods), with the terminal densities of the one-dimensional QMT evolutions in place of the conditionals~\eqref{eq:si-map-conditional}.

\textit{Targets known up to normalization}: If the target is known only as an unnormalized density $\tilde\pi=c\pi$ with unknown $c>0$, as for Boltzmann distributions, no samples are needed. The map can be learned by minimizing the reverse Kullback--Leibler divergence $D_{\mathrm{KL}}(\rho_{\bS}\Vert\pi)=\E_{\rho_{\bS}}[\log(\rho_{\bS}/\tilde\pi)]+\log c$, in which the unknown constant $c$ does not affect the minimizer. For QMT, the model density $\rho_\theta$ (Methods) is available in simulation on the whole detector grid, so the corresponding objective $\int_{\Odet}\rho_\theta\log(\rho_\theta/\tilde\pi)\dd x$ can be evaluated by quadrature. The main text mentions this route for targets given in closed form, but all examples in this work use samples.

\subsection{One-dimensional QMT evolutions as map components}
\label{sec:si-kr-qmt}

\begin{proposition}[Monotone QMT flow]\label{prop:si-monotone}
Let $\psi\in C^1([0,t'];L^2(\R))\cap C([0,t'];H^2(\R))$ be a smooth, normalized solution of equation~\eqref{eq:si-schrodinger} on $\R$ with $\rho=\abs{\psi}^2>0$, let $v=j/\rho$, and let $F_t$ be the CDF of $\rho(\cdot,t)$. Then $X_t=F_t^{-1}\circ F_0$ is a strictly increasing bijection of $\R$ and is the flow of $v$ on $[0,t']$. If, for previously measured coordinates $x_{1:k-1}$, the terminal density is the target conditional $\pi(\cdot\mid x_{1:k-1})$, then $X_{t'}$ is the unique increasing map that pushes the input density $\abs{\psi(\cdot,0)}^2$ forward to this conditional density. It equals $S_k^{-1}(\cdot\,;x_{1:k-1})$ for the Knothe--Rosenblatt map~\eqref{eq:si-kr-cdf-map} with $\Phi$ replaced by the input CDF $F_0$.
\end{proposition}

\begin{proof}
Because $\rho(\cdot,t)>0$ has unit mass, $F_t$ is a smooth, strictly increasing bijection from $\R$ onto $(0,1)$, so $X_t$ is well defined and strictly increasing. Since $\psi(\cdot,t)$ and $\partial_x\psi(\cdot,t)$ lie in $H^1(\R)$, they vanish at infinity, so $j\to0$ as $x\to-\infty$; moreover, $\rho\in C^1([0,t'];L^1(\R))$. The conservation law $\partial_t\rho+\partial_xj=0$ therefore gives $\partial_tF_t(y)=\int_{-\infty}^y\partial_t\rho\,\dd y'=-j(y,t)$, so $X_t(x)$ is continuously differentiable in $t$ by the implicit function theorem ($\partial_yF_t=\rho>0$), and differentiating $F_t(X_t(x))=F_0(x)$ in $t$ yields $\rho(X_t,t)\,\partial_tX_t=j(X_t,t)$, that is, $\partial_tX_t=v(X_t,t)$ with $X_0(x)=x$. Solutions of this equation are unique because $v$ is smooth, so $X_t$ is the flow of $v$.
\end{proof}

Measuring the position at $t'$ is therefore equivalent in distribution to drawing $z_k$ from the input density and applying $X_{t'}$: the quantum measurement replaces the reference draw $z_k$ in equation~\eqref{eq:si-recursive-inverse}, and the flow map $X_{t'}$ takes the role of $S_k^{-1}$. This is the sequential scheme of main-text Fig.~2b. The sampling scheme remains valid even if the wavefunction has zeros (nodes), where $v=j/\rho$ is undefined, because a detection at $t'$ samples $\abs{\psi(\cdot,t')}^2$ in any case.

\subsection{Block-triangular maps in the handwritten-digit experiment}
\label{sec:si-mnist}

In the experiment of main-text Fig.~4b--h, each photon detection on the two-dimensional camera yields two coordinates, so the latent vector is sampled one pair of coordinates at a time (a block-triangular factorization). The digit-conditioned autoencoder (Methods) encoded an image of class $y\in\{0,\ldots,9\}$ as a latent vector $\bx\in\R^{16}$ (denoted $z$ in main-text Fig.~4b,c), and the class-conditioned triangular map $\bS_y$ was fitted to the encoded training images by minimizing equation~\eqref{eq:si-triangular-sample-loss} with the parameterization~\eqref{eq:si-monotone-param} and $r_k(h)=\operatorname{softplus}(h)+10^{-3}$. The factorization~\eqref{eq:si-chain-rule} of the model density, grouped into eight pairs, reads
\begin{equation}
    \rho(\bx\mid y)
    =\rho(x_1,x_2\mid y)
    \prod_{\ell=2}^{8}
    \rho(x_{2\ell-1},x_{2\ell}\mid y,x_{1:2\ell-2}),
    \label{eq:si-mnist-chain}
\end{equation}
and by equation~\eqref{eq:si-map-conditional} each pair density is the product of two map conditionals,
\begin{equation}
    \rho(x_{2\ell-1},x_{2\ell}\mid y,x_{1:2\ell-2})
    =\prod_{k=2\ell-1}^{2\ell}
    \eta_1\!\left(S_{y,k}(x_{1:k})\right)
    \frac{\partial S_{y,k}(x_{1:k})}{\partial x_k},
    \label{eq:si-mnist-pair-density}
\end{equation}
where $x_{1:0}$ is empty. For each pair, this density was evaluated at the points $(\xi_i,\xi_j)$ of a $32\times32$ grid covering a common latent range and normalized to cell probabilities $p_{ij}$ with $\sum_{i,j}p_{ij}=1$. The probability matrix was displayed on the SLM (Supplementary Note~5). An accepted photon detection (Methods) selected a cell $(i,j)$ and set $(x_{2\ell-1},x_{2\ell})=(\xi_i,\xi_j)$, which conditioned all subsequent pairs. Training, evaluation of the conditional densities, the deterministic completion of concentrated conditionals (Methods), and decoding were digital. The random cell selections were made by single-photon detection.

\section{Optical Schr\"odinger equation}
\label{sec:si-optics}

This note derives equation~(3) in the main text and its correspondence with the Schr\"odinger equation of Supplementary Notes~1 and~2. For a monochromatic field $\mathcal{E}$ in the scalar approximation (a single field component of fixed polarization), the source-free Helmholtz equation in one transverse dimension $x$ is\cite{Milonni1992}
\begin{equation}
    \left[\frac{\partial^2}{\partial z^2}+\frac{\partial^2}{\partial x^2}+k_0^2n^2(x,z)\right]
    \mathcal{E}(x,z)=0 .
    \label{eq:si-helmholtz-scalar}
\end{equation}
Assume predominantly forward propagation along $z$ and write
\begin{equation}
    \mathcal{E}(x,z)=E(x,z)\exp(i k_0 n_{\text{ref}} z),
\end{equation}
where $k_0=2\pi/\lambda_0$ is the vacuum wavenumber, $n_{\text{ref}}>0$ a constant reference refractive index close to the values of $n$, and $E$ the field envelope of the main text. Substitution gives
\begin{equation}
    \frac{\partial^2 E}{\partial z^2}+2ik_0n_{\text{ref}}\frac{\partial E}{\partial z}+\frac{\partial^2 E}{\partial x^2}
    +k_0^2\left[n^2(x,z)-n_{\text{ref}}^2\right]E=0 .
\end{equation}
The slowly varying envelope approximation, $\abs{\partial^2 E/\partial z^2}\ll2k_0n_{\text{ref}}\abs{\partial E/\partial z}$, valid for propagation at small angles to the $z$ axis, neglects $\partial^2 E/\partial z^2$ and yields the paraxial equation
\begin{equation}
    i\frac{\partial E}{\partial z}
    =
    -\frac{1}{2k_0n_{\text{ref}}}\frac{\partial^2 E}{\partial x^2}
    -\frac{k_0}{2n_{\text{ref}}}\left[n^2(x,z)-n_{\text{ref}}^2\right]E .
    \label{eq:si-paraxial}
\end{equation}
This equation, sometimes called the optical Schr\"odinger equation, coincides with main-text equation~(1) divided by $\hbar$ under the substitutions
\begin{equation}
    \psi\leftrightarrow E,\qquad
    t\leftrightarrow z,\qquad
    \frac{\hbar}{m}\leftrightarrow\frac{1}{k_0n_{\text{ref}}},\qquad
    \frac{V}{\hbar}\leftrightarrow
    -\frac{k_0}{2n_{\text{ref}}}\left[n^2(x,z)-n_{\text{ref}}^2\right].
    \label{eq:si-optical-potential-map}
\end{equation}
For a beam with two transverse dimensions, as in main-text Fig.~3, $\partial^2/\partial x^2$ is replaced by $\partial^2/\partial x^2+\partial^2/\partial y^2$ and $n$ by $n(x,y,z)$. Regions with $n>n_{\text{ref}}$ act as potential wells. For $n_{\text{ref}}=1$ and $n=1+\Delta n$ with $\abs{\Delta n}\ll1$, the potential term becomes $-k_0\Delta n$ to first order, which is the form of the dimensionless potential $U_\theta$ used for main-text Figs.~1 and~2 (Methods). For real $n$ and a field that is periodic in $x$ or decays as $\abs{x}\to\infty$, equation~\eqref{eq:si-paraxial} conserves $\int\abs{E}^2\dd x$, and the normalized intensity $\abs{E}^2/\int\abs{E}^2\dd x$ is the probability density of the transverse position at which a photon is detected.

\section{Single-layer potentials and the SLM experiment}
\label{SN:single_layer}

In the experiment of Fig.~4 in the main text, the three-dimensional index profile of Fig.~3 is replaced by a spatial light modulator (SLM) placed between crossed polarizers and imaged onto a single-photon camera. This note describes this configuration as a limiting case of QMT and derives the pattern that encodes a target distribution. With two transverse dimensions, equation~(3) in the main text reads
\begin{equation}
i\frac{\partial}{\partial z}E(x,y,z) =
\left[-\frac{1}{2k_0n_{\text{ref}}}\Delta + V(x,y,z)\right]E(x,y,z),
\label{eq:SN_paraxial}
\end{equation}
with $\Delta=\partial^2/\partial x^2+\partial^2/\partial y^2$ and $V(x,y,z) = -(k_0/2n_{\text{ref}})\left[n^2(x,y,z)-n_{\text{ref}}^2\right]$.
A thin element at $z=0$ corresponds to a potential concentrated in a single layer, $V(x,y,z)=g(x,y)\,\delta(z)$. Integrating equation~\eqref{eq:SN_paraxial} across the layer, where the diffraction term does not contribute, gives
\begin{equation}
E(x,y,0^+) = e^{-ig(x,y)}\,E(x,y,0^-)~.
\label{eq:SN_phase}
\end{equation}
For real $g$, the layer imposes a phase $-g(x,y)$ and leaves the intensity unchanged. The target must then be formed by subsequent propagation, and $g$ can be obtained by inverse design. This is the lossless single-layer version of QMT. For $g=-i\Gamma$ with $\Gamma\geq0$, the layer attenuates the field,
\begin{equation}
E(x,y,0^+) = e^{-\Gamma(x,y)}\,E(x,y,0^-)~,
\label{eq:SN_amp}
\end{equation}
and removes a fraction of the photons, so the evolution is nonunitary. A phase-only liquid-crystal SLM between crossed polarizers realizes such amplitude modulation. For input light polarized at $45^\circ$ to the liquid-crystal axis, the analyzer transmits the component rotated by the local retardance $\varphi(x,y)$, with intensity transmission $\sin^2(\varphi/2)$. The transmission amplitude also carries a phase, $g=i\log a$ for a complex amplitude $a(x,y)$, but this phase does not affect the detected intensity because the SLM plane is imaged onto the camera.

In the experiment, the SLM plane is imaged onto the detector, so no propagation takes place between the layer and the detection plane. For an input intensity $I_{\mathrm{in}}(x,y)$, the density of detected photons is
\begin{equation}
\rho(x,y)=\frac{e^{-2\Gamma(x,y)}I_{\mathrm{in}}(x,y)}{\iint e^{-2\Gamma(x',y')}I_{\mathrm{in}}(x',y')\,\dd x'\dd y'}~.
\label{eq:SN_rho}
\end{equation}
It equals $\rho_{\mathrm{tar}}$ if and only if $e^{-2\Gamma}\propto\rho_{\mathrm{tar}}/I_{\mathrm{in}}$ where $I_{\mathrm{in}}>0$. The choice that maximizes the transmitted power under $\Gamma\geq0$ is
\begin{equation}
\Gamma(x,y) = -\frac12\ln\frac{\rho_{\mathrm{tar}}(x,y)/I_{\mathrm{in}}(x,y)}{\max_{x',y'}\left[\rho_{\mathrm{tar}}(x',y')/I_{\mathrm{in}}(x',y')\right]}~,
\label{eq:SN_gamma}
\end{equation}
which reduces to $\Gamma=-\frac12\ln(\rho_{\mathrm{tar}}/\max\rho_{\mathrm{tar}})$ for uniform illumination. The target is thus reached by removing probability rather than by transporting it. With uniform illumination over an area $|A|$, the fraction of photons transmitted is
\begin{equation}
\eta_{\mathrm{ph}}=\frac{1}{|A|\max_{x,y}\rho_{\mathrm{tar}}(x,y)},
\label{eq:SN_efficiency}
\end{equation}
the ratio of the mean to the maximum of the target density. For a distribution over $N$ uniformly illuminated cells of equal area with probabilities $p_j$, $\eta_{\mathrm{ph}}=1/(N\max_jp_j)$. For the latent conditionals of main-text Fig.~4b ($N=32^2$), $\eta_{\mathrm{ph}}$ therefore ranges from a few per cent for broad conditionals to about $10^{-3}$ when one cell carries 90\% of the probability. A real potential followed by propagation, as in main-text Figs.~1--3, reaches the target without these losses.

In the experiment, each distribution is a set of cell probabilities $p_j$: $32\times32$ latent-grid cells for the digits (Supplementary Note~3.4) and 16 patches of $7\times7$ camera pixels for the tokens (Methods). The accepted-event probability of cell $B_j$ is
\begin{equation}
q_j=\frac{\int_{B_j}T(r)I_{\mathrm{in}}(r)\,\dd r}{\sum_{\ell}\int_{B_\ell}T(r)I_{\mathrm{in}}(r)\,\dd r},
\label{eq:SN_cells}
\end{equation}
where $T=e^{-2\Gamma}$ is the intensity transmission after the analyzer, and the sum runs over the accepted cells (`hot pixels' and rejected cells excluded; Methods). The pattern is chosen such that $q_j\approx p_j$. The finite on--off ratio $\kappa\approx100$ (Extended Data Fig.~1b) limits the transmission to $T\geq T_{\max}/\kappa$, so that probabilities below about $p_{\max}/\kappa$ are raised towards this floor. Detections in cells with programmed probability below $10^{-10}$ are rejected (Methods), which removes this leakage for empty cells but not for small nonzero probabilities. Optical blur, detector response, and residual background can further change the relation between $q_j$ and $p_j$. Extended Data Fig.~2 compares programmed, measured, and single-photon distributions.

\section{Target distributions}
\label{sec:si-target-dists}

We draw the target samples for Figs.~1--3 in the main text from the distributions below: the banana, checkerboard, ring, and double-crescent distributions for Fig.~2d,e and, in physical coordinates, for Fig.~3 (Methods); the $x_1$ marginal of the double crescent for Fig.~1c; and the noisy Archimedean spiral for Fig.~1d. We write $X=(X_1,X_2)$ for a random vector and $x=(x_1,x_2)$ for a realization or the argument of a density. Lengths are dimensionless; target samples outside the detector window are clipped coordinatewise to it (Methods).

The centred one-dimensional Gaussian density with standard deviation $\sigma$ and the centred two-dimensional Gaussian density with covariance $\sigma^2I_2$ are
\begin{equation}
\varphi_\sigma(z)
=
\frac{1}{\sqrt{2\pi}\sigma}
\exp\left(
-\frac{z^2}{2\sigma^2}
\right),
\quad
z\in\mathbb R,
\qquad
\varphi_\sigma^{(2)}(z)
=
\frac{1}{2\pi\sigma^2}
\exp\left(
-\frac{\lVert z\rVert_2^2}{2\sigma^2}
\right),
\quad
z\in\mathbb R^2.
\end{equation}

\textit{Banana distribution}: The banana density is
\begin{equation}
\rho_{\mathrm{banana}}(x_1,x_2)
=
\varphi_1(x_1)
\varphi_{0.3}
\left[
x_2-\frac{1}{2}\left(x_1^2-1\right)
\right],
\label{eq:si-banana-density}
\end{equation}
that is, $X_1\sim\mathcal N(0,1)$ and $X_2\mid X_1\sim\mathcal N\big(\frac{1}{2}(X_1^2-1),0.3^2\big)$.

\textit{Checkerboard distribution}: The $4\times4$ checkerboard has cell width $c=0.8$, jitter standard deviation $\sigma_{\mathrm{jit}}=0.05$, and offset $o=(1.6,1.6)$. The set of active cells is
\begin{equation}
\mathcal A
=
\left\{
(p,q)\in\{0,1,2,3\}^2:
p+q\equiv0\pmod 2
\right\},
\qquad
\abs{\mathcal A}=8.
\label{eq:si-checkerboard-cells}
\end{equation}
The checkerboard density is
\begin{equation}
\rho_{\mathrm{checker}}(x)
=
\frac{1}{\abs{\mathcal A}c^2}
\sum_{(p,q)\in\mathcal A}
\int_{[0,c)^2}
\varphi_{\sigma_{\mathrm{jit}}}^{(2)}
\left(
x-\left[(pc,qc)+u-o\right]
\right)
\dd u.
\label{eq:si-checkerboard-density}
\end{equation}
To generate a sample, choose an active cell uniformly at random, draw a point uniformly in it, shift the point by $-o$ and add independent jitter $\mathcal N(0,\sigma_{\mathrm{jit}}^2I_2)$.

\textit{Ring distribution}: Five rings of nominal radius $R_{\mathrm{ring}}=1$ and radial standard deviation $\tau_{\mathrm{ring}}=0.05$ are centred at $\mu_1=(-2,0)$, $\mu_2=(0,0)$, $\mu_3=(2,0)$, $\mu_4=(-1,-1.2)$, and \linebreak $\mu_5=(1,-1.2)$. To generate a sample, draw independently $J\sim\operatorname{Unif}\{1,\ldots,5\}$, $\Theta\sim\operatorname{Unif}[0,2\pi]$, and $\Xi\sim\mathcal N(0,1)$, and set
\begin{equation}
R=R_{\mathrm{ring}}+\tau_{\mathrm{ring}}\Xi,
\qquad
X=\mu_J+R(\cos\Theta,\sin\Theta).
\label{eq:si-ring-sampling}
\end{equation}

\textit{Double-crescent distribution}: Draw independently $B\sim\operatorname{Bernoulli}(1/2)$, $\Theta\sim\operatorname{Unif}[0,\pi]$, and $\Xi\sim\mathcal N(0,1)$, set $R=2+0.2\Xi$, and define
\begin{equation}
X=
\begin{cases}
(R\cos\Theta-1,\;R\sin\Theta), & B=0,\\
(R\cos\Theta+1,\;-R\sin\Theta-0.1), & B=1.
\end{cases}
\end{equation}
Our datasets are stratified: exactly half of the samples have $B=0$ and half $B=1$.

\textit{Noisy Archimedean spiral}: Draw independently $\Theta\sim\operatorname{Unif}[\pi/2,5\pi]$ and $\Xi\sim\mathcal N(0,I_2)$, and set
\begin{equation}
R=0.2\Theta,
\qquad
X
=
(R\cos\Theta,\;R\sin\Theta)
+
0.15\Xi.
\label{eq:si-spiral-sampling}
\end{equation}

\bibliography{refs,SI_refs}

%% file: preamble.tex
\usepackage[T1]{fontenc}
\usepackage[utf8]{inputenc}
\usepackage{amsmath,amssymb,amsfonts,mathtools,bm}
\usepackage{graphicx}
\usepackage{siunitx,pifont,upgreek,ragged2e}
\usepackage{microtype}
\usepackage{needspace}
\usepackage[hidelinks]{hyperref}
\setcitestyle{super,comma,sort&compress}
\makeatletter
\newcounter{EDfigure}
\def\fps@EDfigure{tbp}
\def\ftype@EDfigure{4}
\def\ext@EDfigure{loedf}
\def\fnum@EDfigure{Extended Data Fig.~\theEDfigure}
\def\frontmatter@above@affiliation@script{%
  \skip@\@flushglue
  \@flushglue\z@ plus.3\hsize\relax
  \centering
  \@flushglue\skip@
  \par
  \begingroup
    \frontmatter@affiliationfont
    \textsuperscript{*}These authors contributed equally.\par
  \endgroup
}
\newenvironment{EDfigure}{\@dblfloat{EDfigure}}{\end@dblfloat}
\def\section{\@startsection{section}{1}{\z@}{1.5ex \@plus .5ex}{.75ex}{\normalfont\normalsize\bfseries}}
\def\subsection{\@startsection{subsection}{2}{\z@}{1.2ex \@plus .4ex}{.6ex}{\normalfont\small\bfseries}}
\def\@hangfrom@section#1#2#3{\@hangfrom{#1#2}#3}
\def\@hangfroms@section#1#2{#1#2}

\makeatother

\newenvironment{methods}{\section{Methods}}{}
\newcommand{\methodsubsection}[1]{%
  \par\addvspace{0.45\baselineskip}%
  \noindent{\normalfont\small\bfseries #1}\par\nobreak\vspace{0.2\baselineskip}\noindent
}

\newenvironment{addendum}{\begin{description}\small}{\end{description}}

\newcommand{\dd}{\,\mathrm{d}}

\newcommand{\abs}[1]{\left\lvert #1 \right\rvert}

%% file: extended_data.tex
\newcommand{\EDFigureOne}{%
\begin{EDfigure}[t]
    \centering
    \includegraphics[width=\linewidth]{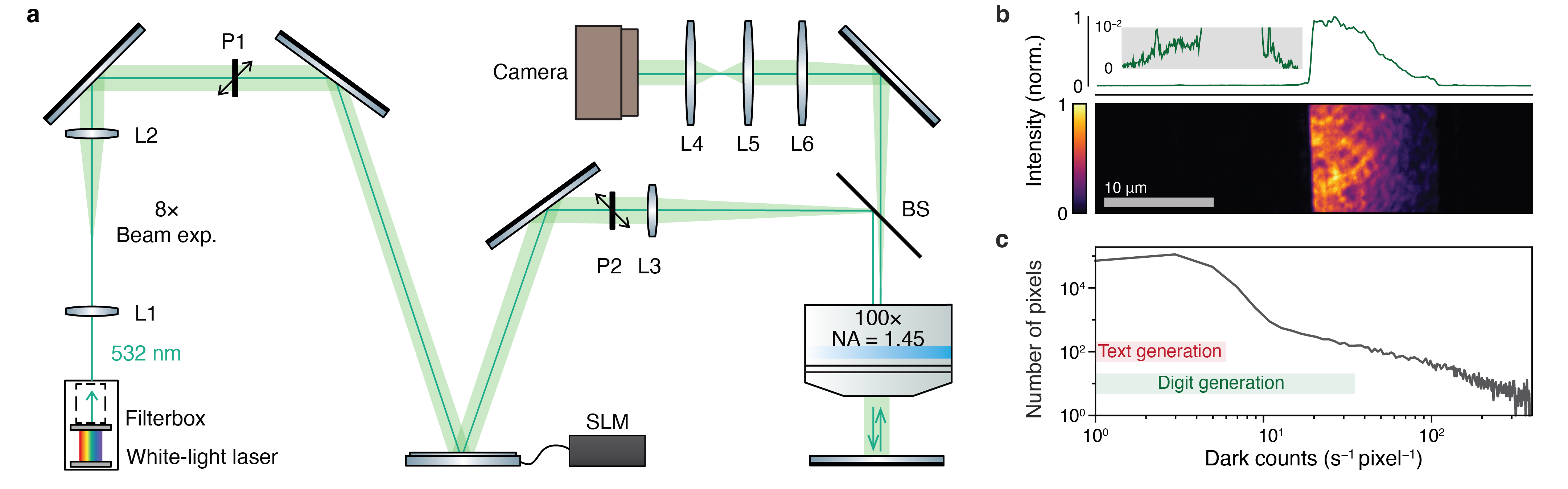}
    \caption{\textbf{Optical setup for physical sampling with single photons.} \textbf{a}, More detailed schematic of the optical setup for single-photon sampling. A telescope (L1 and L2) expands the beam (8×) and the SLM between crossed polarizers modulates the amplitude of the beam to the desired probability distribution. The distribution is mapped onto the mirror on top of the microscope (L3 and the microscope objective) and is further relayed and projected onto the single-photon camera (L4, L5, and L6). See Methods for further details. \textbf{b}, Characterization of the achievable amplitude modulation. One half of the beam (left part of image) was blocked as completely as possible to quantify the dynamic range. The resulting intensity distribution was recorded with a 16-bit EMCCD camera. The inset shows a line trace along the horizontal axis, demonstrating an on--off ratio of at least 100:1. \textbf{c}, Histogram of dark-count rates measured on the single-photon camera. The distribution exhibits a high-count tail arising from `hot pixels'. To suppress their effect on single-photon sampling, pixels with dark count rates exceeding 6 counts s\textsuperscript{$-1$} pixel\textsuperscript{$-1$} were excluded for text-generation experiments, whereas a more permissive threshold of 35 counts s\textsuperscript{$-1$} pixel\textsuperscript{$-1$} was used for digit-generation experiments.}
    \label{EDfig:1}
\end{EDfigure}
}

\newcommand{\EDFigureTwo}{%
\begin{EDfigure}[t]
    \centering
    \includegraphics[width=\linewidth]{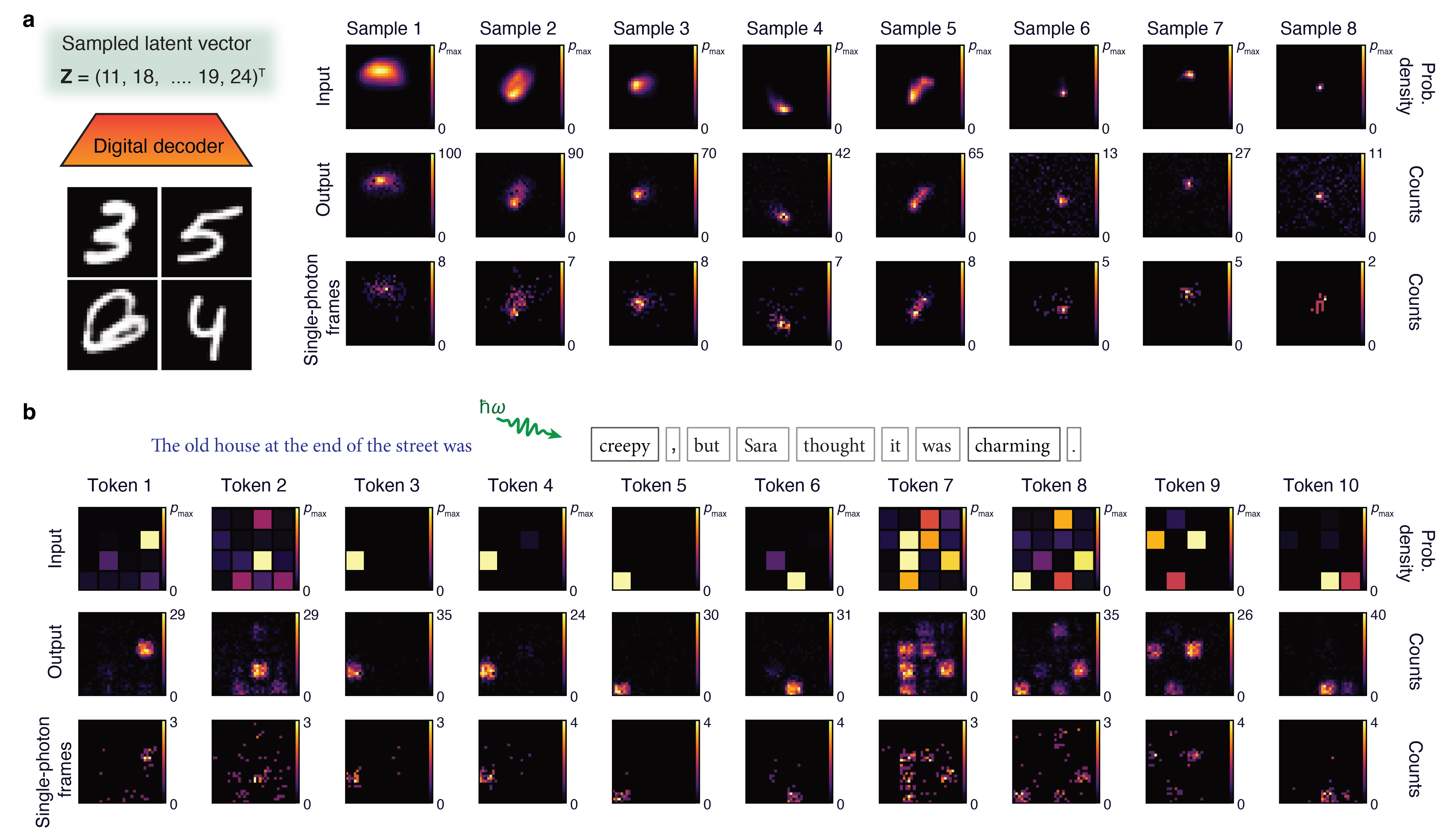}
    \caption{\textbf{Validation of single-photon sampling distributions.} \textbf{a}, Comparison of the input probability distributions (top row), the measured output intensity distributions (middle row, 2\,ms integration time), and the constructed single-photon distributions (bottom row) for the eight samples used in a digit-generation experiment. The single-photon distributions were reconstructed from all frames (500 total) in which exactly one photon was detected. The close agreement between the three distributions demonstrates that the programmed optical probability distributions are reproduced at the single-photon level. \textbf{b}, Same as \textbf{a} but for ten token-probability distributions used in the text-generation experiments, showing similarly close agreement between the input, output, and single-photon distributions.}
    \label{EDfig:2}
\end{EDfigure}
}

\newcommand{\EDFigureThree}{%
\begin{EDfigure}[t]
    \centering
    \includegraphics[width=\linewidth]{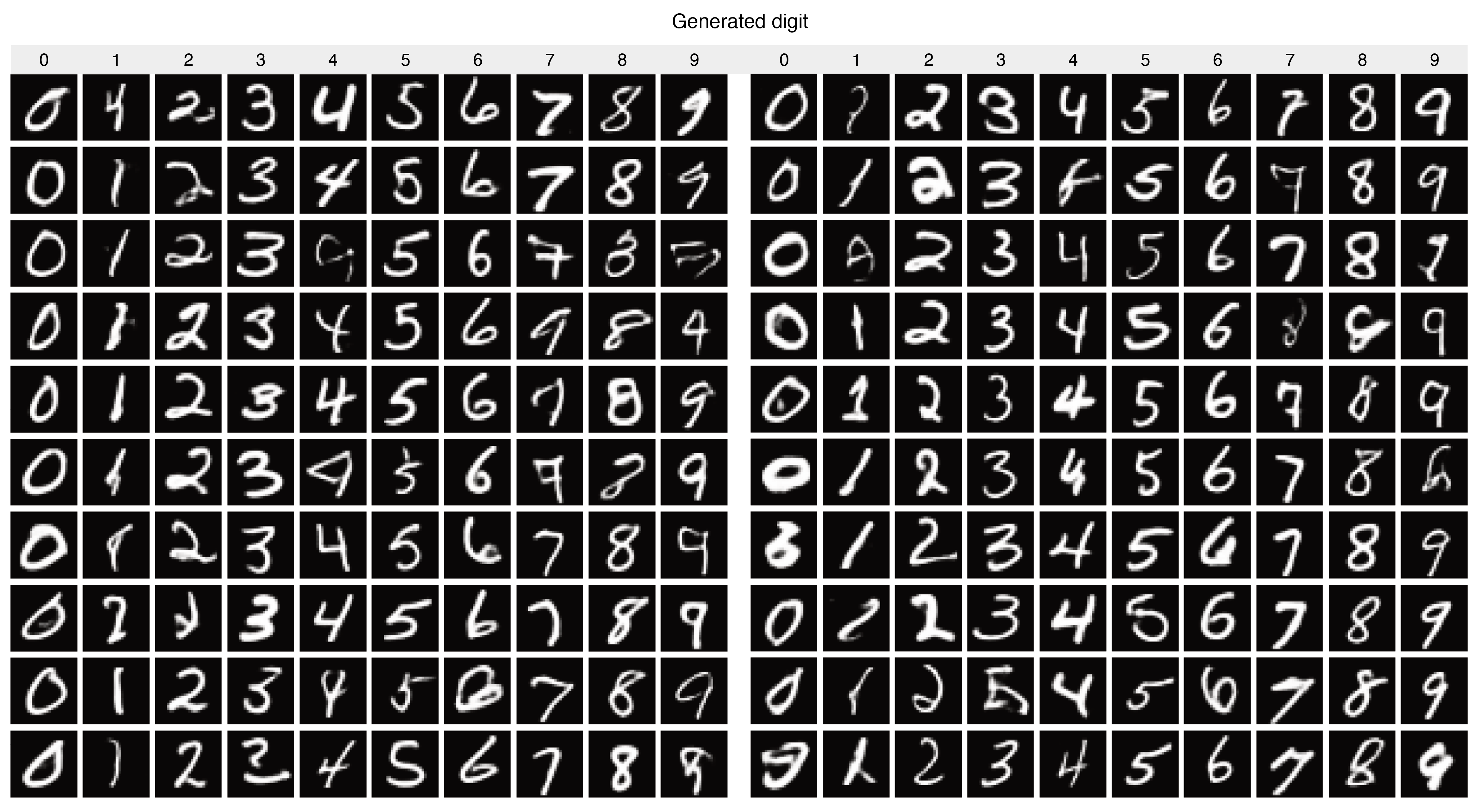}
    \caption{\textbf{All QMT-generated images of handwritten digits.} Two blocks of generated handwritten digits, each arranged as a 10×10 grid. Each column corresponds to one digit class (0--9), with ten independently generated samples per class, together comprising 20 generated samples for each digit class.}
    \label{EDfig:3}
\end{EDfigure}
}

\newcommand{\EDFigureFour}{%
\begin{EDfigure}[t]
    \centering
    \includegraphics[width=\linewidth]{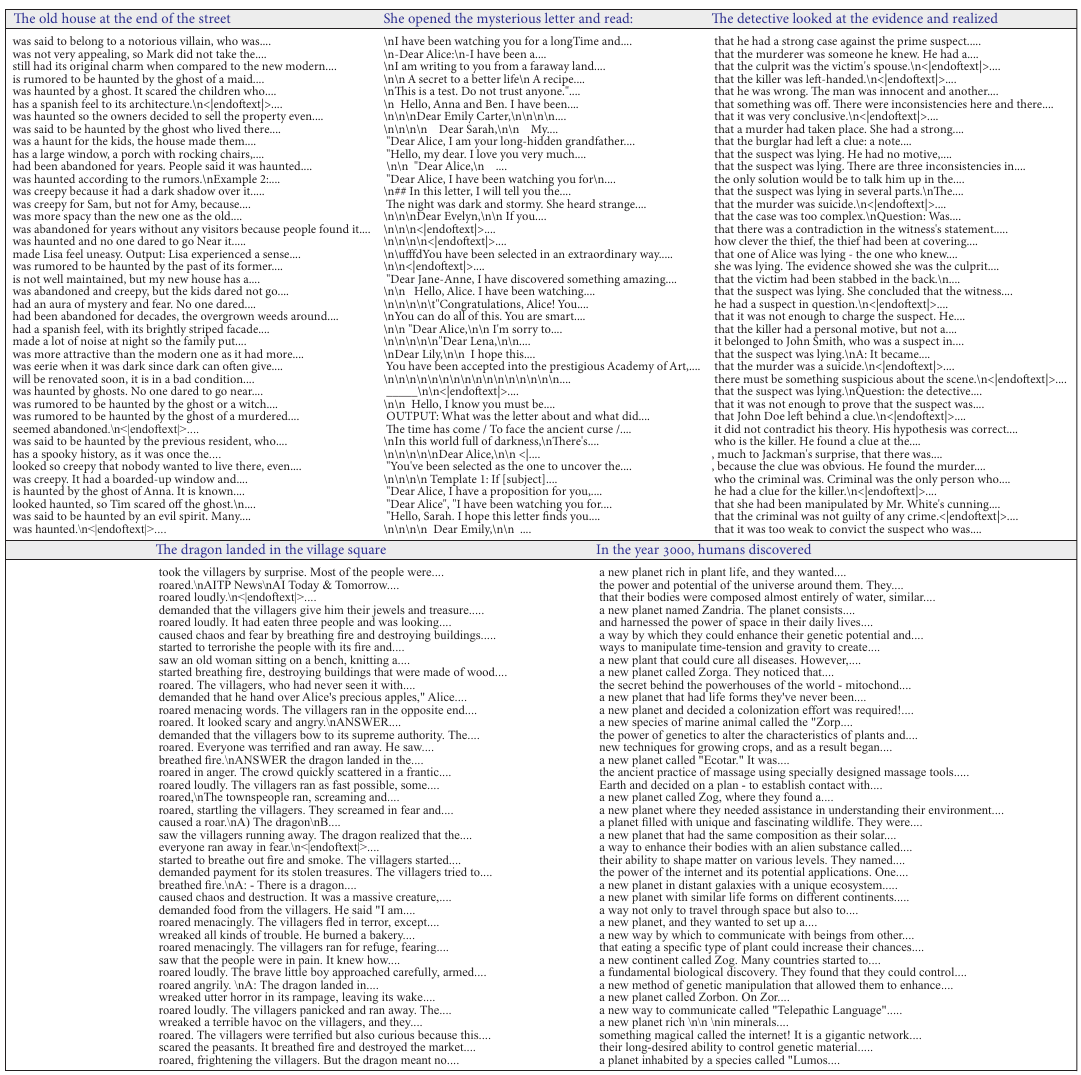}
    \caption{\textbf{All QMT-generated texts.} Further data from the experiment shown in Fig.~\ref{fig:4}. Five different prompts were finished independently 40 times, yielding a total of 200 generated text sequences.}
    \label{EDfig:4}
\end{EDfigure}
}

%% file: methods.tex
\begin{methods}

\methodsubsection{Optimization framework}

The simulations in Figs.~\ref{fig:1}--\ref{fig:3} use a common optimization procedure. A fixed input wavefunction $\psi_0$ is propagated through a parameterized potential, its terminal intensity is normalized over a detector window to obtain a model probability density, and the trainable parameters $\theta$ are optimized by minimizing the negative log-likelihood (NLL) of target samples together with a regularization penalty. Figures~\ref{fig:1} and \ref{fig:2} use a dimensionless propagation model, whereas Fig.~\ref{fig:3} uses optical parameters. For the dimensionless simulations, rescaling equation~\eqref{eq: schrodinger} by characteristic length and time scales $L$ and $T$ gives
\begin{equation}
\begin{aligned}
i\frac{\partial}{\partial t}\psi_\theta(x,t)
&=H_\theta(t)\psi_\theta(x,t),\\
H_\theta(t)&=-A\nabla_x^2+U_\theta(x,t),
\end{aligned}
\label{eq:si-generic-hamiltonian}
\end{equation}
with diffraction coefficient $A=\hbar T/(2mL^2)$ and potential $U_\theta(x,t)=(T/\hbar)V_\theta(Lx,Tt)$. $x$ denotes one transverse coordinate for 1D problems and two for 2D problems. At the terminal propagation coordinate $t'$, the intensity in the detector window $\Omega_{\mathrm{det}}$ gives the model density
\begin{equation}
\rho_\theta(x)
=
\frac{
\abs{\psi_\theta(x,t')}^2
}{
\displaystyle\int_{\Omega_{\mathrm{det}}}
\abs{\psi_\theta(s,t')}^2\dd s
},
\qquad
x\in \Omega_{\mathrm{det}},
\label{eq:si-generic-terminal-density}
\end{equation}
that is, the terminal density conditioned on detection within $\Omega_{\mathrm{det}}$. 

To emulate propagation in unbounded space on a finite grid, the detector window is surrounded by an absorbing layer (a perfectly matched layer in 1D and an absorbing mask in 2D; see below), which removes probability that leaves the window. The evolution is therefore not norm-preserving, and equation~\eqref{eq:si-generic-terminal-density} accounts for the absorbed intensity. The existence results of Supplementary Notes~1 and~2 are proven for norm-preserving evolution on a periodic domain. Supplementary Note~2.3 shows that they also hold for this absorbing, window-conditioned model, and Supplementary Table~S1 summarizes the boundary treatment of all simulations.

We minimize
\begin{equation}
\begin{aligned}
\mathcal L(\theta)
&=\mathcal L_{\mathrm{NLL}}(\theta)+\lambda\mathcal R(\theta),\\
\mathcal L_{\mathrm{NLL}}(\theta)
&=-\frac{1}{N}\sum_{r=1}^{N}\log\rho_\theta(x_r),
\end{aligned}
\label{eq:si-schrodinger-nll}
\end{equation}
where $x_r$ are $N$ target samples in $\Omega_{\mathrm{det}}$, $\mathcal R$ is a regularization term, and $\lambda\geq0$ controls its strength. The NLL is a sample estimate of the cross entropy between the target and model densities, which equals their Kullback--Leibler divergence up to a constant\cite{Cover2006}. For numerical stability, we evaluate the logarithm as $\log[\rho_\theta(x)+10^{-12}]$ and clip target samples outside the detector grid to its nearest endpoint. Gradients are obtained by automatic differentiation through the complete propagation and normalization procedure, and the parameters are updated using Adam\cite{Kingma2015}.

The following sections detail the numerical implementation, training scheme, and parameters used for the simulations in Figs.~\ref{fig:1}--\ref{fig:3}.

\methodsubsection{1D Schr\"odinger simulation (Fig.~\ref{fig:1}c)}

\textit{Solver}: We write the Hamiltonian in equation~\eqref{eq:si-generic-hamiltonian} as
\begin{equation}
\begin{aligned}
H_\theta(t)&=-\frac{1}{2k_0}\frac{\partial^2}{\partial x^2}+U_\theta(x,t),\\
k_0&=2\pi,
\end{aligned}
\label{eq:si-1d-hamiltonian}
\end{equation}
that is, with diffraction coefficient $A=1/(2k_0)$ and trainable potential $U_\theta$. The input field, a centred Gaussian amplitude $\psi_0(x)\propto\exp[-x^2/(2\sigma_0^2)]$, is propagated to $t'$ using the Crank--Nicolson scheme, holding $U_\theta$ fixed during each step $\Delta t$. The computational domain comprises the detector window $[-x_{\mathrm{det}},x_{\mathrm{det}}]$ and a perfectly matched layer (PML) of width $w_{\mathrm{PML}}$ on each side, introduced by replacing $\partial/\partial x$ with $a(x)\partial/\partial x$, where
\begin{equation}
\begin{aligned}
a(x)&=\frac{1}{1+i\sigma(x)/k_0},\\
\sigma(x)&=\gamma_{\mathrm{PML}}\max\bigl(0,\abs{x}-x_{\mathrm{det}}\bigr)^2.
\end{aligned}
\label{eq:si-pml-stretch}
\end{equation}
The resulting operator is discretized on a uniform grid of spacing $h$ using second-order central differences, with $a(x)$ evaluated at the grid points and cell interfaces, and the field is set to zero outside the grid (a homogeneous Dirichlet condition at the outer edge of the PML). Within the PML the discretized Hamiltonian is non-Hermitian, so the Crank--Nicolson update is norm-preserving only in the absence of absorption. The integral in equation~\eqref{eq:si-generic-terminal-density} is evaluated by trapezoidal quadrature. The input field had width $\sigma_0=0.5$ and was propagated to $t'=2$ with step $\Delta t=0.01$ on a grid of spacing $h=0.05$, with a detector half-width $x_{\mathrm{det}}=4$ and PMLs of width $w_{\mathrm{PML}}=2$ and strength $\gamma_{\mathrm{PML}}=10$.

\textit{Potential parameterization}: The potential is expanded in a truncated Fourier basis,
\begin{equation}
U_\theta(x,t)
=
-k_0
\sum_{a,b=1}^{32}
\alpha_{ab}\,
f_a(u_t)\,
f_b(u_x),
\label{eq:si-fourier-dn-1d}
\end{equation}
where $\alpha$ is the trainable coefficient matrix, and $u_t$ and $u_x$ are $t$ and $x$ rescaled to $[0,1]$ over the propagation interval and the full computational domain $[-6,6]$. The basis functions $f_1,\ldots,f_{32}$ are $\sin(k\pi u)$ and $\cos(k\pi u)$ for $k=1,\ldots,16$, with their mean over the grid subtracted.

\textit{Training}: The target is the $x_1$ marginal of the double-crescent distribution (Supplementary Note~6), with $200000$ samples split $80/20$ into training and validation sets. Starting from $\alpha=0$, we train using full batches and a constant learning rate of $10^{-3}$ for $20000$ updates, with loss $\mathcal L_{\mathrm{1D}}=\mathcal L_{\mathrm{NLL}}+10\operatorname{mean}(\alpha^2)$. The coefficients with the lowest validation NLL, evaluated every 50 updates, are retained.

\methodsubsection{2D Schr\"odinger simulation (Fig.~\ref{fig:1}d)}

\textit{Solver}: We extend equation~\eqref{eq:si-1d-hamiltonian} to two transverse coordinates $(x,y)$. The input field, a centred Gaussian amplitude $\psi_0(x,y)\propto\exp[-(x^2+y^2)/(2\sigma_0^2)]$, is propagated to $t'$ with step $\Delta t$ using the symmetric, second-order split-step Fourier update
\begin{equation}
\begin{aligned}
\psi_{m+1}={}&M_{\mathrm{abs}}
\exp\!\Bigl[-\frac{iU_m\Delta t}{2}\Bigr]\\
&\mathcal F^{-1}\!\Bigl\{\exp\!\Bigl[
-\frac{i(k_x^2+k_y^2)\Delta t}{2k_0}\Bigr]\\
&\quad\mathcal F\!\Bigl\{\exp\!\Bigl[-\frac{iU_m\Delta t}{2}\Bigr]
\psi_m\Bigr\}\Bigr\},
\end{aligned}
\label{eq:si-split-step-2d}
\end{equation}
where $U_m=U_\theta(x,y,t_m)$ is held fixed during step $m$, $\mathcal F$ and $\mathcal F^{-1}$ denote the 2D discrete Fourier transform and its inverse, and $k_x$ and $k_y$ are the corresponding angular spatial frequencies. The discrete Fourier transform imposes periodic boundary conditions on the computational grid, which extends beyond the detector window. After each step, the absorbing mask
\begin{equation}
\begin{aligned}
M_{\mathrm{abs}}(x,y)
={}&\exp\bigl\{-\gamma_{\mathrm{abs}}\bigl[\\
&\quad\max\bigl(0,\abs{x}-x_{\mathrm{det}}\bigr)^2\\
&\quad+\max\bigl(0,\abs{y}-x_{\mathrm{det}}\bigr)^2
\bigr]\bigr\}
\end{aligned}
\label{eq:si-2d-absorbing-mask}
\end{equation}
attenuates fields outside the detector window $\Omega_{\mathrm{det}}=[-x_{\mathrm{det}},x_{\mathrm{det}}]^2$ and thereby prevents periodic wrap-around. Applying the mask once per step is a first-order splitting of the evolution with a complex absorbing potential (Supplementary Note~2.3). The integral in equation~\eqref{eq:si-generic-terminal-density} is evaluated by rectangular quadrature. The input field had width $\sigma_0=3$ and was propagated to $t'=2.25$ with step $\Delta t=0.00375$ on a $192\times192$ grid covering $[-5.5,5.5]^2$, with a detector half-width $x_{\mathrm{det}}=4$ and mask strength $\gamma_{\mathrm{abs}}=5$.

\textit{Potential parameterization}: As for the 1D simulation, with a second transverse coordinate,
\begin{equation}
U_\theta(x,y,t)
=
-k_0
\sum_{a,b,c=1}^{12}
\beta_{abc}\,
f_a(u_t)\,
f_b(u_x)\,
f_c(u_y),
\label{eq:si-fourier-dn-2d}
\end{equation}
where $\beta$ is the trainable coefficient tensor, the basis functions $f_1,\ldots,f_{12}$ are defined as above with $k=1,\ldots,6$, and $u_x$ and $u_y$ are rescaled over $[-5.5,5.5]$.

\textit{Training}: The target is the noisy Archimedean spiral (Supplementary Note~6), with the same number of samples and split as for the 1D simulation. Starting from $\beta=0$, we train using full batches and a constant learning rate of $10^{-3}$ for $200000$ updates, with loss $\mathcal L_{\mathrm{2D}}=\mathcal L_{\mathrm{NLL}}+\mathcal R_{\mathrm{2D}}$, where
\begin{equation}
\begin{aligned}
\mathcal R_{\mathrm{2D}}
={}&\frac{0.01}{k_0^2}\Bigl[
\operatorname{mean}\Bigl|\frac{\partial U_\theta}{\partial t}\Bigr|^2\\
&+\operatorname{mean}\Bigl|\frac{\partial U_\theta}{\partial x}\Bigr|^2\\
&+\operatorname{mean}\Bigl|\frac{\partial U_\theta}{\partial y}\Bigr|^2\Bigr]
\end{aligned}
\label{eq:si-2d-control-regularization}
\end{equation}
penalizes spatial and temporal variations in the potential, with derivatives evaluated using forward finite differences. Model selection follows the 1D simulation. In Fig.~\ref{fig:1}d, the potential is displayed as $U_\theta/k_0$, and the target density is a Gaussian kernel density estimate (bandwidth $0.25$) of the spiral samples.

\methodsubsection{Conditional 1D Schr\"odinger simulations (Fig.~\ref{fig:2})}

\textit{Solver}: All propagations use the solver, input field, and numerical parameters of the 1D Schr\"odinger simulation. The targets in Fig.~\ref{fig:2}d,e are the banana, checkerboard, ring, and double-crescent distributions (Supplementary Note~6).

\textit{Potential parameterization}: The first-coordinate marginal is represented by a potential of the form in equation~\eqref{eq:si-fourier-dn-1d}, with its own coefficient matrix $\alpha^{(1)}$. For the second coordinate, a multilayer perceptron $C_\eta$ with two hidden layers of 128 units and hyperbolic-tangent activations takes the conditioning value $x_1$ as input and predicts a correction to the coefficient matrix $\alpha^{(2,0)}$ of an unconditional base potential. The conditional potential $U_\theta^{(2)}(x_2,t\mid x_1)$ is then given by equation~\eqref{eq:si-fourier-dn-1d} with $\alpha$ replaced by $\alpha^{(2)}(x_1)=\alpha^{(2,0)}+C_\eta(x_1)$.

\textit{Training}: Each target dataset contains $200000$ samples. The two marginal models are first trained separately, using an $80/20$ training--validation split, minibatches of 64 samples, and at most $50000$ updates. The learning rate is increased linearly to $10^{-3}$ during the first 500 updates and then decreased to $10^{-6}$ by cosine decay. The loss is $\mathcal L_{\mathrm{NLL}}+\lambda_\alpha\operatorname{mean}(\alpha^2)$, with $\lambda_\alpha=10$ for the first ten updates and thereafter fixed to the ratio of the mean minibatch NLL over these updates to $\operatorname{mean}(\alpha^2)$ after the tenth update, clipped to $[0.1,100]$. Training stops when the validation NLL, evaluated every 200 updates, fails to improve by at least $10^{-4}$ for 20 consecutive evaluations.

The Fourier coefficients and network parameters of the conditional model are then trained on a random subset of $30000$ sample pairs, divided into $24000$ training and $6000$ validation pairs, of which a fixed $4096$ are used to evaluate the validation NLL. This subset may overlap the data used for the marginal models. We minimize $\mathcal L_{\mathrm{cond}}=\mathcal L_{\mathrm{NLL}}+0.1\operatorname{mean}\abs{D_{xx}\rho}^2$, where the NLL is evaluated on the conditional density $\widehat{\rho}_2(x_2\mid x_1)$ and the second term penalizes rapid spatial variations in this density, with $D_{xx}$ the centred second finite difference and the mean taken over minibatch members and interior detector nodes. Training uses minibatches of 256, the same learning-rate schedule, and at most 5000 epochs, and stops when the validation NLL, evaluated every five epochs, fails to improve by at least $10^{-4}$ for 40 consecutive evaluations.

\textit{Sequential generation}: Samples are generated by drawing $x_1$ from $\widehat{\rho}_1$ by inverse cumulative distribution function sampling, evaluating the conditional potential at the sampled $x_1$, and drawing $x_2$ from $\widehat{\rho}_2(\mathord{\cdot}\mid x_1)$ in the same way. For Fig.~\ref{fig:2}d,e, the fitted density $\widehat{\rho}(x_1,x_2)=\widehat{\rho}_1(x_1)\widehat{\rho}_2(x_2\mid x_1)$ is evaluated on a $400\times400$ grid over $[-4,4]^2$.

\methodsubsection{Photonic simulations (Fig.~\ref{fig:3})}

\textit{Solver}: We solve equation~\eqref{eq:optical_schrodinger}, extended to two transverse coordinates $(x,y)$ in physical units, with $n=n_\theta(x,y,z)$ as the trainable refractive-index profile. The input field, a centred Gaussian amplitude $E_0(x,y)\propto\exp[-(x^2+y^2)/(2w_0^2)]$, is propagated over the length $L_z$ using the split-step Fourier update in equation~\eqref{eq:si-split-step-2d}, with $\Delta t$ replaced by the step $\Delta z$, $k_0$ by $k_0n_{\text{ref}}$ in the diffraction step, and $U_m$ by $-k_0(n_m^2-n_{\text{ref}}^2)/(2n_{\text{ref}})$ in the two phase steps, where $k_0=2\pi/\lambda_0$. An absorbing mask surrounding the detector window attenuates outgoing fields to suppress periodic wrap-around. We used a vacuum wavelength $\lambda_0=\SI{1.55}{\micro\meter}$, reference index $n_{\text{ref}}=1.5163$, propagation length $L_z=\SI{220}{\micro\meter}$, and input width $w_0=\SI{30}{\micro\meter}$, on a computational domain $[-\SI{55}{\micro\meter},\SI{55}{\micro\meter}]^2$ with detector window $[-\SI{40}{\micro\meter},\SI{40}{\micro\meter}]^2$.

\textit{Potential parameterization}: The refractive-index profile is constructed from a Fourier expansion of the form of the sum in equation~\eqref{eq:si-fourier-dn-2d}, with $t$ replaced by $z$ and $k=1,\ldots,12$. A smooth bounding function maps this expansion to a profile satisfying $\abs{n_\theta-n_{\text{ref}}}\leq0.005$.

\textit{Training}: The targets are the distributions of Fig.~\ref{fig:2}d,e (Supplementary Note~6) in physical coordinates, with the same number of samples and split as for the 1D simulation.
Optimization is initialized from solutions obtained with six frequency pairs per coordinate. The loss is $\mathcal L_{\mathrm{NLL}}+\mathcal R_{\mathrm{2D}}$, with the smoothness penalty in equation~\eqref{eq:si-2d-control-regularization} applied to $k_0(n_\theta-n_{\text{ref}})$ and $t$ replaced by $z$. Training uses minibatches of $4096$, a learning rate of $10^{-3}$, and gradient clipping at a threshold of $10$, and proceeds in two stages: $3000$ updates on a $48\times48\times114$ grid, followed by $1000$ updates on a $64\times64\times341$ grid. Fields and performance metrics are then evaluated on a $192\times192\times1363$ grid.

\methodsubsection{Optical measurements}

The measurements were performed on a home-built optical setup. A pulsed supercontinuum laser (SuperK FIU-15; NKT Photonics) with an 80 MHz repetition rate provided the light source. The wavelength was set to 532 nm (1.5 nm linewidth) with a tunable filter (LLTF Contrast; NKT Photonics). This was combined with a spatial light modulator (SLM; PLUTO-2.1-NIR-144; Holoeye) and a single-photon sensitive camera (single-photon avalanche diode array, SPAD 512; Pi Imaging).

The entire setup is depicted in Extended Data Fig.~\ref{EDfig:1}a. All optics are from Thorlabs unless noted. The laser beam was expanded by a telescope consisting of a 25-mm lens (L1; LA1951-A) and a 200-mm lens (L2; LA1708-A). The probability distribution was encoded in the optical intensity by placing the SLM between crossed polarizers (P1; LPVISE100-A and P2; WP25M-UB1), enabling on–off intensity modulation with a dynamic range of approximately 100:1 (Extended Data Fig.~\ref{EDfig:1}b). The SLM plane was first imaged onto a silver mirror mounted at the sample plane of an inverted microscope (Ti2-U Eclipse; Nikon) using a 750-mm lens (L3; LA4716-UV) in combination with a 100× objective (CFI Plan Apochromat Lambda; Nikon). The reflected light was collected by the same objective, transmitted through the microscope tube lens (L4; 200-mm tube lens; Nikon) and a beam splitter (BS; BSW26R), and relayed from the microscope image plane onto the single-photon camera using a 200-mm lens (L5; LA1708) and a 300-mm lens (L6; LA1484-AB). The use of the microscope body enabled straightforward imaging of the input probability distributions on a separate EMCCD camera (iXon 888; Andor) for alignment. The overall magnification was chosen such that five SLM pixels (8×8 $\upmu$m$^2$) were mapped onto one single-photon camera pixel (16×16 $\upmu$m$^2$).

\methodsubsection{Handwritten digit generation}

For Fig.~\ref{fig:4}b--h, we trained a digit-conditioned convolutional autoencoder with 16 latent coordinates. With its encoder and decoder fixed, a separate digit-conditioned triangular map was fitted to the encoded training images by minimizing the latent NLL. The map supplied $32\times32$ probability matrices for successive coordinate pairs, conditioned on the digit class and all previously selected coordinates. Accepted photon detections selected pairs. Concentrated conditionals were completed deterministically as described below. The fixed decoder converted the completed latent vector to an image using the same digit class.

For each distribution, 500 frames with an exposure time of 1 $\upmu$s were recorded in 1-bit mode, providing information on whether a pixel detected at least one photon or none. The photon flux was kept sufficiently low to ensure that the probability of recording more than one photon per pixel per frame was negligible and that each detection event corresponded to a single sample. The first single-photon frame was used for selecting components of the latent vector, whereas the remaining single-photon frames were used to verify that the measured single-photon distribution reproduced the programmed probability distribution (Extended Data Fig.~\ref{EDfig:2}). Camera pixels exhibiting high dark-count rates (\textgreater 35 counts s$^{-1}$ pixel$^{-1}$; `hot pixels') were excluded from the analysis (Extended Data Fig.~\ref{EDfig:1}c). Moreover, samples with input probability density below $10^{-10}$ were rejected, as these sampling events likely originated from background light. Physical sampling of the conditional distribution was terminated once more than 90\% of the probability was concentrated in a single pixel. Because all subsequent conditional distributions likewise remained concentrated (\textgreater90\%) in a single pixel, the remaining samples were assigned deterministically to the most probable pixel.

\methodsubsection{Text generation}

The optical setup was interfaced with the Microsoft Phi-2 large language model\cite{Javaheripi2023}. The 16 highest-probability next tokens were encoded onto the SLM, with each token assigned to a distinct 7×7 pixel region on the single-photon camera. For each token, 500 frames with an exposure time of 200\,ns were recorded in 1-bit mode and the first single-photon frame was picked. Again, pixels with high dark-count rates (\textgreater 6 counts s$^{-1}$ pixel$^{-1}$) were excluded. To suppress crosstalk between neighboring token regions due to diffraction, a one-pixel-wide overflow region was inserted between adjacent token patches.

\end{methods}